# A Doubly Robust Estimator for the Mann Whitney Wilcoxon Rank Sum Test When Applied for Causal Inference in Observational Studies

Ruohui Chen[a,b], Tuo Lin[a], Lin Liu[a], Jinyuan Liu[a], Ruifeng Chen[a], Jingjing Zou[a], Chenyu Liu[a], Loki Natarajan[a], Tang Wang[c], Xinlian Zhang[a] and Xin Tu[a]

[a]Division of Biostatistics and Bioinformatics, Herbert Wertheim School of Public Health and Human Longevity Science, University of California, San Diego, La Jolla, California, 92093, USA; [b]Division of Biostatistics, Feinberg School of Medicine, Northwestern University, Chicago, Illinois , 60611, USA; [c]Biostatistics and Data Science, School of Public Health and Tropical Medicine, Tulane University, New Orleans, LA, 70112, USA

**ABSTRACT**
The Mann-Whitney-Wilcoxon rank sum test (MWWRST) is a widely used method for comparing two treatment groups in randomized control trials, particularly when dealing with highly skewed data. However, when applied to observational study data, the MWWRST often yields invalid results for causal inference. To address this limitation, Wu et al. (2014) introduced an approach that incorporates inverse probability weighting (IPW) into this rank-based statistic to mitigate confounding effects. Subsequently, Mao (2018), Zhang et al. (2019), and Ai et al. (2020) extended this IPW estimator to develop doubly robust estimators.

Nevertheless, each of these approaches has notable limitations. Mao's method imposes stringent assumptions that may not align with real-world study data. Zhang et al.'s (2019) estimators rely on bootstrap inference, which suffers from computational inefficiency and lacks known asymptotic properties. Meanwhile, Ai et al. (2020) primarily focus on testing the null hypothesis of equal distributions between two groups, which is a more stringent assumption that may not be well-suited to the primary practical application of MWWRST.

In this paper, we aim to address these limitations by leveraging functional response models (FRM) to develop doubly robust estimators. We demonstrate the performance of our proposed approach using both simulated and real study data.

**KEYWORDS**
Functional Response Models, Inverse Probability Weighting, Mean Score Imputation, Mean Rank, Outcome Regression, U-statistics Generalized Estimating Equations

## 1. Introduction

In randomized control trials, the non-parametric Mann-Whitney-Wilcoxon rank sum test (MWWRST) is widely used as an alternative to the two-sample t-test when data distributions are highly skewed, especially with outliers. For non-randomized studies, this rank-based test generally yields invalid results for causal inference. Although one may remove or winsorize outliers and apply mean-based methods such as regression, propensity score matching, and marginal structure models, results from such analyses are difficult to interpret as they are subjective and depend on how the outliers are handled.

CONTACT Ruohui Chen Email: chenruohui2013@gmail.com

For non-randomized observational studies, Wu et al.(2014)[1] introduced an approach for causal inference by incorporating inverse probability weighting (IPW) into the MWWRST. Their approach addressed limitations of an earlier attempt by Rosenbaum (2002) [2], in which a constant individual treatment effect $\tau = y_1 - y_0$ between two potential outcomes $(y_1, y_0)$ is imposed in order to use a randomization technique for inference. This assumption is not only implausible conceptually, but also unverifiable in practice. Recently, Mao et al. (2018) [3], Zhang et al. (2019) [4] and Ai et al. (2020) [5] extended the IPW approach of Wu et al. (2014) to develop doubly robust estimators for more robust inference. However, their doubly robust estimators have major limitations.

In Mao (2018), parametric logistic and linear regression were used for the propensity score of the IPW and outcome model of the augmented component of the doubly robust estimator. Since the logistic regression cannot address over-dispersion [6, 7], it may not be correct for some real studies. The parametric linear regression is even more problematic, since it not only fails to address outliers, which calls for the MWWRST in the first place, but also limits its applications to normal data. Thus, both parametric models, especially the second one, can be wrong for most real study data and as such this doubly robust estimator may even be less robust than the IPW estimator in Wu et al. (2014), which uses a semiparametric generalized linear model with the logit link, or restricted moment [8], for the propensity score.

Zhang et al. (2019) presented two doubly robust estimators. One estimator also posits a parametric outcome model, which like Mao's approach does not address outliers, since no reliable estimator of this outcome model can be obtained in the presence of outliers. The second estimator addresses the limitation of Mao's parametric outcome model by using a semiparametric regression for between-subject attributes, or generalized probability index model (GPI) as so termed in the paper, since it generalizes the probability index model (PI) developed by Thas et al. (2012) [9]. A major limitation of their approach is the reliance on bootstrap for inference. First, applications of bootstrap within the current context are highly inefficient, since the MWWRST statistic based on a sample size $n$ requires computational times in the order of $O\left(n^2\right)$, where $n = n_1 + n_2$ and $n_k$ denotes the sample size of group $k$ (= 0,1). For example, in a similar application involving modeling between-subject attributes for microbiome beta diversity outcomes [10],the run time for asymptotic inference is about 35 times less compared to a permutation-based approach with 1,000 permutations. For the simulation set up in their paper, we find that the run time for their bootstrap inference is generally around 40 times the run time for the proposed doubly robust estimator in this paper with asymptotic inference. Second, the bootstrap procedure for their doubly robust estimators has an unknown performance even for large samples, since it is being applied to quite a complex U-statistics setting with both estimated model parameters for the propensity score and outcome regression models. As no investigation of large sample properties were conducted, applications of their doubly robust estimators raise questions about inference validity when applied to real study data.

Ai et al. (2020) focused on the semiparametric efficiency of the doubly robust estimator when applied to non-randomized observational studies for causal treatment effects. While deriving asymptotic variances and consistent estimators is an essential theoretical concern, it's also vital to note that Ai et al.'s (2020) focus is on testing the null hypothesis of equal distribution, which differs from the null hypothesis of mean rank equality. As a result, Ai et al.'s (2020) method has more limited applications compared to the approach introduced in the current manuscript [13]. As previously mentioned, the MWWRST becomes necessary when the application of two-sample ttests, whether assuming equal variances or not, is problematic due to susceptibility to outliers. In such scenarios, the focus is on testing the equality of central measures (e.g., mean or mean rank) between two groups, rather than comparing their distributions, as the latter assumption is considerably more stringent. [13].

In this paper, we develop an approach to address all the aforementioned limitations by leveraging the functional response models (FRM) to provide easy-to-compute, reliable, and robust causal inference when applying the Mann-Whitney-Wilcoxon ranksum test to observational study data. While Zhang et al. (2019) and Ai et al. (2020) both employed the probability index model, a component of the FRM, they did not fully harness the potential of the FRM, including the U-statistics-based generalized estimating equations (UGEE). Consequently, they did not provide asymptotic variances for their respective methodologies. Without these asymptotic variances and associated estimators, the practical applicability of these methods to real studies is limited. Furthermore, although Zhang et al. (2019) suggested using bootstrapping for inference, they did not provide a robust justification for the validity of bootstrapping in the context of a complex U-statistic that incorporates estimated model parameters for both the propensity score and outcome regression models.

In Section 2, we review potential outcomes and causal effects for the MWWRST. In Section 3, we present IPW estimators, outcome regression (mean score imputation) estimators, and doubly robust estimators by combining the IPW and outcome regression estimators. In Section 4, we discuss joint inference for all three estimators by leveraging the FRM and U-statistics-based generalized estimating equations. In Section 5, we examine the performances of the doubly robust estimator for small and large samples through both simulated and real study data. We discuss future directions in Section 6.

## 2. Causal Effect for Mann-Whitney-Wilcoxon Rank-sum Test

There are two equivalent forms of the MWWRST [11]. We use the U-statistics expression, also known as the Mann–Whitney form, for the development below unless stated otherwise.

Consider two independent groups, indexed by $k$ (= 0,1), with sample size $n_k$, and let $y_{i_k k}$ denote an observed continuous outcome from the $i_k$ subject in group $k$. The MWWRST is given by:

Mann–Whitney form of MWWRST : $$\widehat{\delta}_n = \frac{1}{n_1}\frac{1}{n_0}\sum_{i_1=1}^{n_1}\sum_{j_0=1}^{n_0} I\left(y_{i_1 1} \leq y_{j_0 0}\right). \quad (1)$$

The U-statistic $\widehat{\delta}_n$ is an unbiased estimator of $\tilde{\delta} = E\left[I\left(y_{i_1 1} \leq y_{j_0 0}\right)\right]$[11]. The MWWRST tests whether the mean rank of $y_{i_k k}$ is the same between the two groups, since the null $H_0 : \tilde{\delta} = \frac{1}{2}$ holds true if and only if $y_{i_k k}$ has the same mean rank [12]. Only under some assumptions does equal mean rank imply equal median [13]. This correct interpretation is important, since there is a long-standing misconception that the MWWRST always compare medians of two distributions.

To apply the concept of potential outcomes to non-randomized studies, consider a sample of size $n$, and let $z_i$ (= 0,1) denote an indicator of treatment assignment , or exposure status, and let $(y_{i1}, y_{i0})$ denote the potential outcomes corresponding to the two treatment conditions $k$ (= 0,1). Then, for a randomized controlled trial (RCT), $y_{ik} \perp z_i$ and we observe one of the potential outcome $y_{i1}$, denoted as $y_{i_1 1}$, or $y_{i0}$, denoted as $y_{i_0 0}$, depending on whether the subject is randomized to group $k = 1$ or $k = 0$ ($1 \le i_k \le n_k$, $n = n_0 + n_1$). As in causal inference about treatment effects based on comparing the means of $y_{ik}$, we would like to use the following to indicate treatment effect:

$$\Delta = E\left[I\left(y_{i1} \le y_{i0}\right)\right], \tag{2}$$

and then apply the counterfactual consistency assumption to estimate Δ using the observed $y_{i_k k}$ [1]. Unfortunately, such an approach does not work.

First, unlike mean-based models, it is impossible to estimate the rank-based Δ using observed $y_{i_k k}$, because only one of the potential outcomes for each subject is observed. Second, $\Delta \neq E\left[I\left(y_{i1} \le y_{j0}\right)\right]$, which is easily seen by noting the fact that the withinsubject pair $(y_{i1}, y_{i0})$ generally has smaller variability than the between-subject pair $(y_{i1}, y_{j0})$. The purpose of defining Δ is to show that we cannot define causal effect for rank-based models using potential outcomes from the same subject as in meanbased models for within-subject attributes such as the marginal structural models. This is because it is impossible to estimate Δ within the current context of rank-based statistics even for RCTs.

Therefore we define causal effects for the MWWRST by:

$$\delta = E\left[I\left(y_{i1} \le y_{j0}\right)\right], \quad \text{for any } (i,j) \in C_2^n, \tag{3}$$

where $C_q^n$ denotes the set of $\binom{n}{q}$ combinations of $q$ distinct elements $(i_1, i_2, \ldots i_q)$ from the integer set $\{1, \ldots, n\}$. For RCTs, $\delta$ is equal to $\tilde{\delta}$ and can be consistently estimated by $\widehat{\delta}_n$ in (1) based on the observed $(y_{i_1 1}, y_{j_0 0})$ $(i_1 \in C_1^{n_1},\ j_0 \in C_1^{n_0})$, which is a subset of the potential outcomes $(y_{i1}, y_{j0})$ $((i,j) \in C_2^n)$. For non-RCTs, $\delta$ is not equal $\tilde{\delta}$, and $\widehat{\delta}_n$ is no longer a consistent estimator of $\delta$. Although $\delta$ cannot be modeled by conventional mean-based regression models for within-subject attributes such as the generalized linear models (either parametric or semiparametric), it can be modeled by the functional response models (FRM) for between-subject attributes, which will be elaborated below.

Note that as MWWRST is usually called for dealing with outliers, which may also occur for count outcomes. When ties are present, the null in this case may be expressed as $H_0 : E\left(I\left(y_{i1} < y_{j0}\right)\right) + \frac{1}{2}E\left(I\left(y_{i1} = y_{j0}\right)\right) = \frac{1}{2}$ [1]. Thus, for count outcomes, we redefine $f_{\mathrm{i}}$ in (4) with $I\left(y_{i1} \le y_{j0}\right)$ replaced by $I\left(y_{i1} < y_{j0}\right) + \frac{1}{2}I\left(y_{i1} = y_{j0}\right)$. Without loss of generality, we focus on (unbounded) continuous $y_{ik}$ unless stated otherwise.

## 3. Doubly Robust Estimator for Mann-Whitney-Wilcoxon Rank-sum Test

Our doubly robust estimator consists of two components, IPW and outcome regression. We first introduce the two components individually, then we show a joint inference of combining the two parts to construct a double robust estimator.

*3.1. Inverse Probability Weighting Estimator*

Let $p = E(z_i)$. For RCTs, $p$ is a known constant indicating the probability of treatment assignment (exposure status). Let

$$f_{\mathbf{i}} = \frac{1}{2}\left(\frac{z_i(1-z_j)}{p(1-p)} I(y_{i1} \leq y_{j0}) + \frac{z_j(1-z_i)}{p(1-p)} I(y_{j1} \leq y_{i0})\right), \quad \mathbf{i} = (i,j) \in C_2^n. \tag{4}$$

Although only one of the potential outcomes $(y_{i1}, y_{i0})$ is observed, $f_{\mathbf{i}}$ in the above is well-defined, since $z_i(1-z_j) = 0$ for pairs of subjects who are assigned the same treatment.

For an RCT, $\widehat{\delta}_n$ in (1) is a consistent estimator of $\delta$ and can be expressed as:

$$\begin{aligned}\widehat{\delta}_n &= (n_1 n_0)^{-1} \sum_{i_1=1}^{n_1} \sum_{j_0=1}^{n_0} I(y_{i_1 1} \leq y_{j_0 0}) \\ &= \left[\frac{1}{2}\sum_{\mathbf{i}\in C_2^n}\left(\frac{z_i(1-z_j)}{p(1-p)} + \frac{z_j(1-z_i)}{p(1-p)}\right)\right]^{-1}\left(\sum_{\mathbf{i}\in C_2^n} f_{\mathbf{i}}\right) \\ &= A_n^{-1}\left(\sum_{\mathbf{i}\in C_2^n} f_{\mathbf{i}}\right),\end{aligned} \tag{5}$$

where $A_n^{-1} = \left[\frac{1}{2}\sum_{\mathbf{i}\in C_2^n}\left(\frac{z_i(1-z_j)}{p(1-p)} + \frac{z_j(1-z_i)}{p(1-p)}\right)\right]^{-1}$. By the theory of U-statistics [11],

$$\delta = E(f_{\mathbf{i}}), \quad \sqrt{n}\binom{n}{2}^{-1}\sum_{\mathbf{i}\in C_2^n}(f_{\mathbf{i}} - \delta) \to_d N\left(0, \sigma_\delta^2\right), \quad \binom{n}{2}^{-1} A_n \to_p 1, \tag{6}$$

where $\to_d$ ($\to_p$) denotes convergence in distribution (probability) and $\sigma_\delta^2$ denotes the asymptotic variance of $\sqrt{n}\binom{n}{2}^{-1}\sum_{\mathbf{i}\in C_2^n}(f_{\mathbf{i}} - \delta)$. It follows from Slutsky's theorem that $\widehat{\delta}_n$ has the same asymptotic distribution as $\binom{n}{2}^{-1}\sum_{\mathbf{i}\in C_2^n} f_{\mathbf{i}}$.

For observational studies, $z_i$ generally depends on $y_{ik}$, and as a result, $\widehat{\delta}_n$ in (5) is generally a biased estimator of $\delta$. Suppose there exists a vector of confounders, or covariates for subject $i$, $\mathbf{w}_i$, such that $y_{ik} \perp z_i \mid \mathbf{w}_i$ ($k = 0,1$). Let $\pi_i = E(z_i \mid \mathbf{w}_i)$, and $\mathbf{w}_{\mathbf{i}} = \{\mathbf{w}_i, \mathbf{w}_j\}$, where $\mathbf{i}$ refers to pairs of $i$ and $j$. We define an inverse probability weighted (IPW) estimator in this case by:

$$f_{\mathbf{i}}^{IPW} = \frac{1}{2}\left(\frac{z_i(1-z_j)}{\pi_i(1-\pi_j)} I(y_{i1} \leq y_{j0}) + \frac{z_j(1-z_i)}{\pi_j(1-\pi_i)} I(y_{j1} \leq y_{i0})\right), \quad \mathbf{i} = (i,j) \in C_2^n. \tag{7}$$

Then based on the law of iterated conditional expectation we have:

$$
\begin{aligned}
E\left(f_{\mathbf{i}}^{IPW}\right) &= \frac{1}{2} E\left[\frac{z_i\left(1-z_j\right)}{\pi_i\left(1-\pi_j\right)} I\left(y_{i1} \leq y_{j0}\right)+\frac{z_j\left(1-z_i\right)}{\pi_j\left(1-\pi_i\right)} I\left(y_{j1} \leq y_{i0}\right)\right] \\
&= \frac{1}{2} E\left[\frac{E[I\left(y_{i1} \leq y_{j0}\right) \mid \mathbf{w}_{\mathbf{i}}]}{\pi_i\left(1-\pi_j\right)} E\left(z_i\left(1-z_j\right) \mid \mathbf{w}_{\mathbf{i}}\right)\right] \\
&+ \frac{1}{2} E\left[\frac{E[I\left(y_{j1} \leq y_{i0}\right) \mid \mathbf{w}_{\mathbf{i}}]}{\pi_j\left(1-\pi_i\right)} E\left(z_j\left(1-z_i\right) \mid \mathbf{w}_{\mathbf{i}}\right)\right] \\
&= \delta .
\end{aligned}
$$

If $\pi_i$ is known, then as in the case of RCTs $\widehat{\delta}_n^{IPW}$ below is a consistent and asymptotically normal estimator of $\delta$:

$$
\widehat{\delta}_n^{IPW}=\binom{n}{2}^{-1} \sum_{\mathbf{i} \in C_2^n} f_{\mathbf{i}}^{IPW} \rightarrow_d \delta, \quad \sqrt{n}\left(\widehat{\delta}_n^{IPW}-\delta\right) \rightarrow_d N\left(0, \sigma_\delta^2\right) \quad, \tag{8}
$$

where $\sigma_\delta^2$ denotes the asymptotic variance.

In practice, $\pi_i$ is unknown and can be modeled using any parametric [3, 14, 15] or semiparametric GLM with a link for a binary response such as the logit link [1, 4]. Let $\pi_i = \pi\left(\mathbf{w}_i;\eta\right)$ with $\eta$ indicating an unknown vector of parameters. We can estimate $\eta$ using maximum likelihood or estimating equations. Let $\widehat{\eta}_n$ denote such an estimator of $\eta$. In this case, $\widehat{\delta}_n^{IPW}$ in (8) will be a function of $\widehat{\eta}_n$, i.e., $\widehat{\delta}_n^{IPW}\left(\widehat{\eta}_n\right)$. By using a Taylor series expansion, we can obtain the asymptotic variance of $\widehat{\delta}_n^{IPW}\left(\widehat{\eta}_n\right)$ that accounts for sampling variability of $\widehat{\eta}_n$ for inference about $\delta$ [1, 3, 14, 15].

### *3.2. Outcome Regression Estimator*

The IPW in Section 3.1 only uses the observed pairs $\left(y_{i_1 1}, y_{j_0 0}\right)$ $\left(i_1 \in C_1^{n_1},\ j_0 \in C_1^{n_0}\right)$. Alternatively, we can posit a model to relate outcome $y_{ik}$ with $\mathbf{w}_i$ and use it to impute missing $y_{ik}$.

This approach was considered by Mao (2018) and Zhang et al. (2019). However, as noted in Section 1, parametric models are at odds with the reason for employing the MWWRST in the first place. Moreover, in all real study applications, no reliable estimator can be obtained for such a parametric model because of outliers.

Zhang et al. (2019) introduced another estimator by positing a semiparametric generalized probability index (GPI) model:

$$
E\left(I\left(y_{i1} \leq y_{j0}\right) \mid \mathbf{w}_{\mathbf{i}}\right) = g\left(\mathbf{w}_{\mathbf{i}};\gamma\right), \qquad \mathbf{i} = (i, j) \in C_2^n, \tag{9}
$$

where $\gamma$ denotes a vector of parameters. The GPI above is a member of functional response models (FRM) for between-subject attributes (see Section 4 for more details about FRM). This particular form of FRM has been used in a similar context to address outliers for linear regression in Chen et al. (2014)[16] and Chen et al. (2016) [17]. We will refer to the GPI in (9) as an FRM in the following discussion unless stated otherwise. The FRM in (9) is much more robust as it only models the conditional mean of $I\left(y_{i1} \leq y_{j0}\right)$ given $\mathbf{w}_{\mathbf{i}}$. In addition, it allows us to directly impute missing $I\left(y_{i1} \leq y_{j0}\right)$ for unobserved $y_{i1}$ or $y_{j0}$.

If $\gamma$ is known, then under $y_{ik} \perp z_i \mid \mathbf{w}_i$ and $y_{jk} \perp z_j \mid \mathbf{w}_j$, with $\mathbf{w}_i$ and $\mathbf{w}_j$ indicate vectors of covariates for different subject $i$ and $j$, we can impute missing $I(\{y_{i1} \leq y_{j0}\})$ or $I(\{y_{j1} \leq y_{i0}\})$ with the mean score (MS), $g(\mathbf{w}_{\mathbf{i}};\gamma)$ or $g(\mathbf{w}_{\mathbf{i}^c};\gamma)$ with $\mathbf{w}_{\mathbf{i}} = \{\mathbf{w}_i, \mathbf{w}_j\}$ and $\mathbf{w}_{\mathbf{i}^c} = \{\mathbf{w}_j, \mathbf{w}_i\}$. For example, under the logit link and additive linear predictor, we can posit:

$$g(\mathbf{w}_{\mathbf{i}};\gamma) = \text{expit}\left(\gamma_0 + \boldsymbol{\gamma}_{11}^\top \mathbf{w}_i + \boldsymbol{\gamma}_{10}^\top \mathbf{w}_j\right), \quad g(\mathbf{w}_{\mathbf{i}^c};\gamma) = \text{expit}\left(\gamma_0 + \boldsymbol{\gamma}_{11}^\top \mathbf{w}_j + \boldsymbol{\gamma}_{10}^\top \mathbf{w}_i\right).$$

where expit$(\cdot)$ denotes the inverse of the logit link.

Note that $g(\mathbf{w}_{\mathbf{i}};\gamma)$ is the mean of the semiparametric FRM for between-subject attributes $\{I(y_{i1} \leq y_{j0}), \mathbf{w}_{\mathbf{i}}\}$ in (9), which generally does not have a direct relationship with semiparametric GLM for within-subject attributes $\{y_{ik}, \mathbf{w}_i\}$. For example, if conditioning on $\mathbf{w}_i$, $y_{ik}$ follows a semiparametric linear regression:

$$y_{ik} = \beta_0 + \boldsymbol{\beta}_1^\top \mathbf{w}_{ik} + \epsilon_{ik}, \quad 1 \leq i \leq n_k, \quad k = 0, 1.$$

then the link function $g^{-1}$ for the semiparametric FRM in (9) is determined by the distribution of $\epsilon_{ij} = \epsilon_{i1} - \epsilon_{j0}$. For normal-distributed $\epsilon_{ik}$, $\epsilon_{ij}$ is also normal and $g^{-1}$ is the probit link:

$$g(\mathbf{w}_{\mathbf{i}};\gamma) = \Phi(\gamma_0 + \boldsymbol{\gamma}_{11}^\top \mathbf{w}_i + \boldsymbol{\gamma}_{10}^\top \mathbf{w}_j), \tag{10}$$

where $\Phi(\cdot)$ denotes the cumulative distribution function (CDF) of the standard normal $N(0,1)$.

The following $f_{\mathbf{i}}^{MSI}$ based on such mean score imputed (MSI) $g(\mathbf{w}_{\mathbf{i}};\gamma)$ and $g(\mathbf{w}_{\mathbf{i}^c};\gamma)$ for the unobserved $I(y_{i1} \leq y_{j0})$ and $I(y_{j1} \leq y_{i0})$ is well-defined for all $\binom{n}{2}$ subject pairs of the combined sample:

$$\begin{aligned} f_{\mathbf{i}}^{MSI} = & \frac{1}{2}\left[z_i(1-z_j) I(y_{i1} \leq y_{j0}) + (1 - z_i(1-z_j)) g(\mathbf{w}_{\mathbf{i}};\gamma)\right] \\ & + \frac{1}{2}\left[z_j(1-z_i) I(y_{j1} \leq y_{i0}) + (1 - z_j(1-z_i)) g(\mathbf{w}_{\mathbf{i}^c};\gamma)\right]. \end{aligned} \tag{11}$$

Also, we have:

$$\begin{aligned} E\left(f_{\mathbf{i}}^{MSI}\right) &= \frac{1}{2} E\left[z_i(1-z_j) I(y_{i1} \leq y_{j0}) + (1 - z_i(1-z_j)) g(\mathbf{w}_{\mathbf{i}};\gamma)\right] \\ &\quad + \frac{1}{2} E\left[z_j(1-z_i) I(y_{j1} \leq y_{i0}) + (1 - z_j(1-z_i)) g(\mathbf{w}_{\mathbf{i}^c};\gamma)\right] \\ &= \frac{1}{2}\left[E\left\{z_i(1-z_j) E\left(I(y_{i1} \leq y_{j0}) - g(\mathbf{w}_{\mathbf{i}};\gamma)\right) \mid \mathbf{w}_{\mathbf{i}}, z_i, z_j\right\} + E\left[g(\mathbf{w}_{\mathbf{i}};\gamma)\right]\right] \\ &\quad + \frac{1}{2}\left[E\left\{z_j(1-z_i) E\left(I(y_{j1} \leq y_{i0}) - g(\mathbf{w}_{\mathbf{i}^c};\gamma)\right) \mid \mathbf{w}_{\mathbf{i}^c}, z_i, z_j\right\} + E\left[g(\mathbf{w}_{\mathbf{i}^c};\gamma)\right]\right] \\ &= \frac{1}{2} E\left[g(\mathbf{w}_{\mathbf{i}};\gamma)\right] + \frac{1}{2} E\left[g(\mathbf{w}_{\mathbf{i}^c};\gamma)\right] \\ &= \delta. \end{aligned}$$

Thus by the theory of U-statistics, the estimator $\widehat{\delta}_n^{MSI}$ based on the mean score imputed $f_{\mathbf{i}}^{MSI}$ is consistent and asymptotically normal with asymptotic variance $\tau_\delta^2$:

$$\widehat{\delta}_n^{MSI} = \binom{n}{2}^{-1} \sum_{\mathbf{i} \in C_2^n} f_{\mathbf{i}}^{MSI} \to_d \delta, \quad \sqrt{n}\left(\widehat{\delta}_n^{MSI} - \delta\right) \to_d N\left(0, \tau_\delta^2\right).$$

If $\gamma$ is unknown as in most applications, we can estimate $\gamma$ using U-statistics based generalized estimating equations for semiparametric FRM (see Section 4 for details) [4, 14]. As in the case of IPW estimators above, we can also account for sampling variability in estimated $\widehat{\gamma}_n$ in the asymptotic variance of $\widehat{\delta}_n^{MSI}(\widehat{\gamma}_n)$ [16, 17]. This asymptotic inference addresses the limitation of bootstrap inference proposed by Zhang et al. (2019).

### *3.3. Doubly Robust Estimator*

First, we assume $\eta$ and $\gamma$ are known and discuss how to construct a doubly robust estimator of $\delta$ by combining the two.

To this end, let

$$\begin{aligned} f_{\mathbf{i}}^{DR} &= \frac{1}{2}\left[\frac{z_i(1-z_j)}{\pi_i(1-\pi_j)} I(y_{i1} \le y_{j0}) + \left(1 - \frac{z_i(1-z_j)}{\pi_i(1-\pi_j)}\right) g_{\mathbf{i}}\right] \\ &\quad + \frac{1}{2}\left[\frac{z_j(1-z_i)}{\pi_j(1-\pi_i)} I(y_{j1} \le y_{i0}) + \left(1 - \frac{z_j(1-z_i)}{\pi_j(1-\pi_i)}\right) g_{\mathbf{i}^c}\right], \\ \pi_i &= \pi(\mathbf{w}_i;\eta), \quad g_{\mathbf{i}} = g(\mathbf{w}_{\mathbf{i}};\gamma), \quad g_{\mathbf{i}^c} = g(\mathbf{w}_{\mathbf{i}^c};\gamma), \quad \mathbf{i} = (i,j) \in C_2^n. \end{aligned} \tag{12}$$

The above $f_{\mathbf{i}}^{DR}$ is well-defined for all $\binom{n}{2}$ subject pairs of the combined sasmple. We now show that $E\left(f_{\mathbf{i}}^{DR}\right) = \delta$, if one of the regression models, $\pi(\mathbf{w}_i;\eta)$ and $g(\mathbf{w}_{\mathbf{i}};\gamma)$, is correctly specified. Since the two terms in $f_{\mathbf{i}}^{DR}$ have the same mean, it suffices to show this result only for the first term, i.e.,

$$\begin{aligned} E(I_{\mathbf{i}}) &= E\left[\frac{z_i(1-z_j)}{\pi_i(1-\pi_j)} I(y_{i1} \le y_{j0}) + \left(1 - \frac{z_i(1-z_j)}{\pi_i(1-\pi_j)}\right) g_{\mathbf{i}}\right] \\ &= \delta. \end{aligned}$$

1. If $E(I(y_{i1} \le y_{j0}) \mid \mathbf{w}_{\mathbf{i}}) = g(\mathbf{w}_{\mathbf{i}};\gamma)$ is correctly specified, then we have:

$$\begin{aligned} E(I_{\mathbf{i}}) &= E(I(y_{i1} \le y_{j0})) + E\left[\frac{z_i(1-z_j) - \pi_i(1-\pi_j)}{\pi_i(1-\pi_j)} \left(I(y_{i1} \le y_{j0}) - g(\mathbf{w}_{\mathbf{i}};\gamma)\right)\right] \\ &= E(I(y_{i1} \le y_{j0})) + E\left\{E\left[\frac{z_i(1-z_j) - \pi_i(1-\pi_j)}{\pi_i(1-\pi_j)} \left(I(y_{i1} \le y_{j0}) - g(\mathbf{w}_{\mathbf{i}};\gamma)\right) \mid \mathbf{w}_{\mathbf{i}}, z_i, z_j\right]\right\} \\ &= E(I(y_{i1} \le y_{j0})) + E\left\{\frac{z_i(1-z_j) - \pi_i(1-\pi_j)}{\pi_i(1-\pi_j)} \left[g(\mathbf{w}_{\mathbf{i}};\gamma) - g(\mathbf{w}_{\mathbf{i}};\gamma)\right]\right\} \\ &= \delta. \end{aligned}$$

2. If $E(z_i \mid \mathbf{w}_{\mathbf{i}}) = \pi(\mathbf{w}_i;\eta)$ is correctly specified, then we have:

$$
\begin{aligned}
E(I_{\mathbf{i}}) &= E\left[\frac{z_i(1-z_j)}{\pi_i(1-\pi_j)} I(y_{i1} \le y_{j0}) + \left(1 - \frac{z_i(1-z_j)}{\pi_i(1-\pi_j)}\right) g(\mathbf{w_i};\gamma)\right] \\
&= E\left\{E\left[\frac{z_i(1-z_j)}{\pi_i(1-\pi_j)} I(y_{i1} \le y_{j0})\right] \mid \mathbf{w_i}, y_{i1}, y_{j0}\right\} \\
&+ E\left\{E\left[\left(1 - \frac{z_i(1-z_j)}{\pi_i(1-\pi_j)}\right) g(\mathbf{w_i};\gamma)\right] \mid \mathbf{w_i}\right\} \\
&= E\left[\frac{I(y_{i1} \le y_{j0})}{\pi_i(1-\pi_j)} E(z_i(1-z_j) \mid \mathbf{w_i})\right] \\
&+ E\left[g(\mathbf{w_i};\gamma)\left(1 - \frac{1}{\pi_i(1-\pi_j)} E(z_i(1-z_j) \mid \mathbf{w_i})\right)\right] \\
&= \delta.
\end{aligned}
$$

Thus, if $\pi(\mathbf{w}_i;\eta)$ or $g(\mathbf{w_i};\gamma)$ is correctly specified, it follows from the theory of Ustatistics that the estimator $\delta_n^{DR}$b below based on $f_{\mathbf{i}}^{DR}$ in (12) is consistent and asymptotically normal with asymptotic variance $\varsigma_\delta^2$:

$$
\widehat{\delta}_n^{DR} = \binom{n}{2}^{-1} \sum_{\mathbf{i} \in C_2^n} f_{\mathbf{i}}^{DR} \to_d \delta, \quad \sqrt{n}\left(\widehat{\delta}_n^{DR} - \delta\right) \to_d N\left(0, \varsigma_\delta^2\right)
$$
.

Since $\eta$ and $\gamma$ are both unknown as in most applications, we may again first estimate $\eta$ and $\gamma$ using maximum likelihood (as in Mao, 2018 and Zhang et al, 2019) or U-statistics based generalized estimating equations through semiparametric GLM and FRM (as in Wu et al., 2014 and Chen et al., 2016). Given such estimators, we then compute $\widehat{\delta}_n^{DR}(\widehat{\boldsymbol{\eta}}_n, \widehat{\gamma}_n)$ and its asymptotic variance estimates. When modeling the outcome regression using FRM, the asymptotic inference again addresses the limitation of bootstrap inference in Zhang et al. (2019). Alternatively, we may utilize the flexibility of FRM to jointly estimate $\delta$, $\eta$ and $\gamma$, as we discuss next.

## 4. Inference for Functional Response Models

We first provide a brief overview of the functional response models (FRM). More details can be found in Kowalski and Tu (2007) [11] and other citations below.

Let $y_i$ and $\mathbf{x}_i$ denote some response and a vector of predictors (or covariates) from the $i$th subject ($1 \le i \le n$). The class of functional response model (FRM) is defined by:

$$
E\left[f\left(y_{i_1}, \ldots, y_{i_q}; \boldsymbol{\theta}\right) \mid \mathbf{x}_{i_1}, \ldots, \mathbf{x}_{i_q}\right] = h\left(\mathbf{x}_{i_1}, \ldots, \mathbf{x}_{i_q}; \boldsymbol{\theta}\right), \quad (i_1, \ldots, i_q) \in C_q^n, \qquad (13)
$$

where $f(\cdot)$ is some function, $h(\cdot)$ is some smooth function (e.g., continuous secondorder derivatives), $C_q^n$ denotes the set of $\binom{n}{q}$ combinations of $q$ ($\ge 1$) distinct elements $(i_1,\ldots,i_q)$ from the integer set $\{1,\ldots,n\}$ and $\vartheta$ a vector of parameters. For $q = 1$, $f(y_i;\vartheta) = y_i$ and $h(\mathbf{x}_i;\boldsymbol{\theta}) = h(\boldsymbol{\theta}^\top \mathbf{x}_i)$, (13) reduces to the semiparametric, or restricted moment, generalized linear models: $E(y_i \mid \mathbf{x}_{i_1}) = h(\boldsymbol{\theta}^\top \mathbf{x}_i),\ 1 \le i \le n$. For $q = 2$, $f(y_i, y_j;\theta) = I(y_i \le y_j)$ and $h(\mathbf{x}_i, \mathbf{x}_j;\theta) = E[I(y_i \le y_j)] = \theta$, (13) reduces to

(3) for causal effects of the MWWRST. By extending traditional semiparametric regression models for within-subject attributes to between-subject attributes, FRM has been utilized to model between-subject relationships in a wide range of applications such as social network connectivity [18], Beta-diversity in microbiome research [10],and a host of popular reliability indices such as Pearson and concordance correlation coefficients [14, 15, 19, 20]. Within the current context, we utilize FRM to facilitate inference when applying the doubly robust estimator to observational study data. We focus on joint inference about $\boldsymbol{\theta} = \left(\delta, \boldsymbol{\eta}^\top, \boldsymbol{\gamma}^\top\right)^\top$ for the doubly robsut estimator $\widehat{\delta}_n^{DR}$.

Similar FRMs are readily developed for joint inference about the parameters for the IPW and MSI estimators.

Consider an FRM of the form:

$$E\left(\mathbf{f_i} \mid \mathbf{w_i}\right) = \mathbf{h_i}\left(\mathbf{w_i}; \vartheta\right), \qquad \mathbf{i} = (i,j) \in C_2^n \tag{14}$$

$$\mathbf{f_i} = (f_{\mathbf{i}1}, f_{\mathbf{i}2}, f_{\mathbf{i}3})^\top, \qquad \mathbf{h_i} = (h_{\mathbf{i}1}, h_{\mathbf{i}2}, h_{\mathbf{i}3})^\top, \qquad \mathbf{w_i} = \{\mathbf{w}_i, \mathbf{w}_j\}$$

$$f_{\mathbf{i}1} = \frac{1}{2}(z_i + z_j), \quad f_{\mathbf{i}2} = \frac{1}{2}\left[I(y_{i1} \le y_{j0}) + I(y_{j1} \le y_{i0})\right]$$

$$f_{\mathbf{i}3} = \frac{1}{2}\left[\frac{z_i(1-z_j)}{\pi_i(1-\pi_j)} I(y_{i1} \le y_{j0}) + \left(1 - \frac{z_i(1-z_j)}{\pi_i(1-\pi_j)}\right) g_{\mathbf{i}}\right] + \frac{1}{2}\left[\frac{z_j(1-z_i)}{\pi_j(1-\pi_i)} I(y_{j1} \le y_{i0}) + \left(1 - \frac{z_j(1-z_i)}{\pi_j(1-\pi_i)}\right) g_{\mathbf{i}^c}\right]$$

$$h_{\mathbf{i}1}\left(\mathbf{w_i}; \boldsymbol{\theta}\right) = \frac{1}{2}\left(\pi\left(\mathbf{w}_i; \boldsymbol{\eta}\right) + \pi\left(\mathbf{w}_j; \boldsymbol{\eta}\right)\right),$$

$$h_{\mathbf{i}2}\left(\mathbf{w_i}; \boldsymbol{\theta}\right) = \frac{1}{2}\left[g\left(\mathbf{w_i}, \boldsymbol{\gamma}\right) + g\left(\mathbf{w}_{\mathbf{i}^c}, \boldsymbol{\gamma}\right)\right], \quad h_{\mathbf{i}3}\left(\mathbf{w_i}; \boldsymbol{\theta}\right) = \delta$$

$$\pi_i = \pi\left(\mathbf{w}_i; \boldsymbol{\eta}\right), \quad \pi_j = \pi\left(\mathbf{w}_j; \boldsymbol{\eta}\right), \quad g_{\mathbf{i}} = g\left(\mathbf{w_i}, \boldsymbol{\gamma}\right), \quad g_{\mathbf{i}^c} = g\left(\mathbf{w}_{\mathbf{i}^c}, \boldsymbol{\gamma}\right)$$

$$\boldsymbol{\theta} = \left(\boldsymbol{\eta}^\top, \boldsymbol{\gamma}^\top, \delta\right)^\top, \quad \boldsymbol{\eta} = \left(\eta_0, \boldsymbol{\eta}_1^\top\right)^\top, \quad \boldsymbol{\gamma} = \left(\gamma_0, \boldsymbol{\gamma}_{11}^\top, \boldsymbol{\gamma}_{10}^\top\right)^\top$$

Let

$$S_{\mathbf{i}} = \mathbf{f_i} - \mathbf{h_i}, \quad D_{\mathbf{i}} = \frac{\partial}{\partial \boldsymbol{\theta}} \mathbf{h_i}\left(\boldsymbol{\theta}\right), \quad M_{\mathbf{i}} = \frac{\partial}{\partial \boldsymbol{\theta}^\top} \mathbf{f_i}\left(\boldsymbol{\theta}\right) - \frac{\partial}{\partial \boldsymbol{\theta}^\top} \mathbf{h_i}\left(\boldsymbol{\theta}\right), \qquad \mathbf{i} = (i,j) \in C_2^n.$$

$$\mathbf{V_i} = Var\left(\mathbf{f_i} \mid \mathbf{w_i}\right) = \begin{pmatrix} V_{\mathbf{i}1} & 0 & 0 \\ 0 & V_{\mathbf{i}2} & 0 \\ 0 & 0 & V_{\mathbf{i}3} \end{pmatrix}^{\frac{1}{2}} R\left(\boldsymbol{\alpha}\right) \begin{pmatrix} V_{\mathbf{i}1} & 0 & 0 \\ 0 & V_{\mathbf{i}2} & 0 \\ 0 & 0 & V_{\mathbf{i}3} \end{pmatrix}^{\frac{1}{2}}, \tag{15}$$

$$V_{\mathbf{i}1} = Var\left(f_{\mathbf{i}1} \mid \mathbf{w_i}\right), \qquad V_{\mathbf{i}2} = Var\left(f_{\mathbf{i}2} \mid \mathbf{w_i}\right), \qquad V_{\mathbf{i}3} = Var\left(f_{\mathbf{i}3} \mid \mathbf{w_i}\right)$$

where $\mathbf{V_i}$ is used as the working covariance for the functional response of $\mathbf{f_i}$ given $\mathbf{w_i}$, and has three components $V_{\mathbf{i}1}$, $V_{\mathbf{i}2}$ and $V_{\mathbf{i}3}$, corresponding to the three components of the functional $\mathbf{f_i}$. All three components of $\mathbf{V_i}$ are readily evaluated (see Appendix) and $R(\alpha)$ denotes a working correlation matrix. Inference about $\vartheta$ is based on a class of U-statistics-based generalized estimating equations (UGEE):

$$\mathbf{U}_n\left(\vartheta\right) = \mathrm{X}\, \mathbf{U}_{n,\mathbf{i}} = \mathrm{X}\, D_{\mathbf{i}} V_{\mathbf{i}}^{-1} S_{\mathbf{i}} = \mathbf{0}. \tag{16}$$

$\mathbf{i}\in C_2^n$ $\mathbf{i}\in C_2^n$

If either $\pi(\mathbf{w}_i;\eta)$ or $g(\mathbf{w}_\mathbf{i},\gamma)$ is correctly specified, then we have:

$$E\left(\mathbf{U}_{n,\mathbf{i}}\right)=E\left(D_\mathbf{i}V_\mathbf{i}^{-1}S_\mathbf{i}\right)=E\left[E\left(D_\mathbf{i}V_\mathbf{i}^{-1}S_\mathbf{i}\mid \mathbf{w}_\mathbf{i}\right)\right]=\mathbf{0}.$$

Thus the UGEE in (16) is unbiased, yielding consistent estimators of $\vartheta$, if either $\pi(\mathbf{w}_i;\eta)$ or $g(\mathbf{w}_\mathbf{i},\gamma)$ is correctly specified. Further, UGEE estimators $\widehat{\vartheta}$ are also asymptotically normal under mild regularity conditions. We summarize the asymptotic properties in a theorem below for ease of reference, with details of a proof, along with the exact regularity conditions, given in the Appendix.

**Theorem 1.** Let

$$\mathbf{v}_i=E\left(\mathbf{U}_{n,\mathbf{i}}\mid y_{i1},y_{i0},z_i,\mathbf{w}_i\right),\quad \Sigma=Var\left(\mathbf{v}_i\right),\quad B=E\left(D_\mathbf{i}V_\mathbf{i}^{-1}M_\mathbf{i}^\top\right). \tag{17}$$

Then, under mild regularity conditions, we have:

1. $\widehat{\vartheta}$ is consistent.
2. If $\sqrt{n}(\widehat{\alpha}-\alpha)=\mathbf{O}_p(1)$, i.e., $\widehat{\alpha}$ is $\sqrt{n}$-consistent [11], $\widehat{\vartheta}$ is asymptotically normal:

$$\sqrt{n}\left(\widehat{\boldsymbol{\theta}}-\boldsymbol{\theta}\right)\to_d N\left(\mathbf{0},\Sigma_\theta=4B^{-1}\Sigma B^{-\top}\right). \tag{18}$$

To estimate $\Sigma_\theta$, we note that $\Sigma=Var\left(\mathbf{v}_i\right)=E\left(\mathbf{v}_i\mathbf{v}_i^\top\right)$. Thus,

$$\widehat{\Sigma}=\frac{1}{n}\sum_{i=1}^{n}\widehat{\mathbf{v}}_i\widehat{\mathbf{v}}_i^\top,\quad \widehat{\mathbf{v}}_i=\frac{1}{n-1}\sum_{j\neq i}^{n}\mathbf{U}_{n,ij}. \tag{19}$$

Also, we estimate $B$ by $\widehat{B}=\binom{n}{2}^{-1}\sum_{(i,j)\in C_2^n}\widehat{D}_\mathbf{i}\widehat{V}_\mathbf{i}^{-1}\widehat{M_\mathbf{i}}^\top$, where $\widehat{A}_\mathbf{i}$ denotes $A_\mathbf{i}$ with $\widehat{\vartheta}$ substituting for $\vartheta$ in $A_\mathbf{i}$. Thus, a consistent estimate of $\Sigma_\theta$ is given by: $\widehat{\Sigma}_\theta=4\widehat{B}^{-1}\widehat{\Sigma}\widehat{B}^{-\top}$.

## 5. Application

We illustrate the proposed approach with both simulated and real data. We start with investigating the performances of the doubly robust estimator for small and large samples by simulated data and then present an application to a real weight-loss trial to improve physical activities for breast cancer survivors. In all examples, we set a two-sided type I $\alpha=0.05$. All analyses are carried out using codes developed under the R software platform [21].

### *5.1. Simulation Study*

In order to investigate the causal effect between two treatment groups under confounding bias, we generate data from the following setup for the potential outcomes, confounders, and treatment assignment mechanisms:

$$y_{ik} = \beta_0 + \beta_1 I\,(z_i = k) + \beta_2 w_i + b_i + \epsilon_{ik}, \tag{20}$$

logit(

$$\epsilon_{ik} \sim (\chi_1^2 - 1)\sqrt{\frac{\sigma^2}{2}}, \quad b_i \sim (\chi_1^2 - 1)\sqrt{\frac{\sigma_b^2}{2}},$$
$$w_i \sim N(\mu_w, \sigma_w^2), \quad z_i \sim Bern(\pi\,(w_i; \boldsymbol{\eta})), \quad k = 0, 1, \quad 1 \leq i \leq n,$$
$$E(z_i \mid w_i)) = \eta_0 + \eta_1 w_i, \quad \pi\,(w_i; \eta) = \frac{\exp\,(\eta_0 + \eta_1 w_i)}{1 + \exp\,(\eta_0 + \eta_1 w_i)},$$

where $z_i = 1$ denotes the treated condition, $Bern(\pi_i)$ denotes a Bernoulli distribution with mean $\pi_i$, $w_i$ is a baseline covariate connecting $z_i$ and $y_{ik}$, and $y_{ik}$ is the potential outcome under the influence of $w_i$, i.e., $\mathbf{y}_i \perp z_i \mid \mathbf{w}_i$. The amount of confounding bias for $\mathbf{y}_i$ is controlled through $\eta_1$. It follows from (20) that:

$$\begin{aligned}
y_{i1} &= \beta_0 + \beta_1 I\,(z_i = 1) + \beta_2 w_i + b_i + \epsilon_{i1} \\
&= \beta_0 + \beta_1 + \beta_2 w_i + b_i + \epsilon_{i1} \\
y_{j0} &= \beta_0 + \beta_1 I\,(z_j = 0) + \beta_2 w_j + b_j + \epsilon_{j0} \\
&= \beta_0 + 0 + \beta_2 w_j + b_j + \epsilon_{j0} \\
E[I((y_{i1} - y_{j0}) \leq 0 \mid \mathbf{w_i})] &= P((\epsilon_{i1} + b_i) - (\epsilon_{j0} + b_j) \leq -\beta_2(w_i - w_j) - \beta_1 \mid \mathbf{w_i}) \\
&= \Phi(\gamma_0 + \gamma_{11}^\top w_i + \gamma_{10}^\top w_j)
\end{aligned}$$
$$\gamma_0 = -\frac{1}{\sqrt{2(\sigma^2 + \sigma_b^2)}}\beta_1, \quad \gamma_{11}^\top = -\frac{1}{\sqrt{2(\sigma^2 + \sigma_b^2)}}\beta_2, \quad \gamma_{10}^\top = \frac{1}{\sqrt{2(\sigma^2 + \sigma_b^2)}}\beta_2$$

where $\Phi(\cdot)$ is the cumulative distribution function (CDF) of the standard normal with a mean of 0 and a standard deviation of 1.

Parameters for the simulation study are set to:

$$\boldsymbol{\beta} = (\beta_0, \beta_1, \beta_2)^\top = (0, 0, 1), \qquad \eta = (\eta_0, \eta_1) = (1, -1),$$
$$\sigma^2 = \sigma_b^2 = 1, \qquad u_w = 1, \qquad \sigma_w^2 = 0.25$$

The true treatment effect is controlled through $\beta_1$, which is set to 0. With a negative $\eta_1$ and positive $\beta_2$, the potential outcomes $y_{ik}$ have a smaller mean in the treated group ($z_i = 1$).

Shown in Table 2 are estimated $\vartheta$ for the FRM models for the three causal estimators, $\widehat{\delta}_{IPW}$, $\widehat{\delta}_{MSI}$ and $\widehat{\delta}_{DR}$, along with their asymptotic and empirical standard errors for the different sample sizes under 1,000 Monte Carlo samples. Table 2 also includes estimates $\widehat{\delta}_{MMW}$ from the standard MWWRST in (1) based on observed outcome $y_{ikk}$, which were set to the observed potential outcome $y_{ik}$ corresponding to the value of $z_i$, i.e., $y_{i1}$ ($y_{i0}$) if $z_i = 1$ (0) with $n_1 = \sum_{i=1}^{n} z_i$ and $n_0 = n - n_1$. The estimates of $\eta$ and standard errors for $\widehat{\delta}_{IPW}$ and estimates of $\gamma$ and standard errors for $\widehat{\delta}_{MSI}$ were all close to their counterparts for $\widehat{\delta}_{DR}$ across all sample sizes. All three causal estimates were quite close to the true $\delta$ = 0.5 and the empirical type I errors were close to the nominal $\alpha$ = 0.05. In contrast, estimates of $\widehat{\delta}_{MMW}$ exhibited high bias. The asymptotic standard errors of the causal estimators were nearly identical to their empirical counterparts. The DR estimator $\widehat{\delta}_{DR}$ had the smaller standard errors across all sample sizes, demonstrating the higher efficiency of $\widehat{\delta}_{DR}$ over $\widehat{\delta}_{IPW}$ and $\widehat{\delta}_{MSI}$, except for $n$ = 50, in which case the asymptotic standard error of $\widehat{\delta}_{DR}$ was slightly larger.

To demonstrate the doubly robust properties of $\widehat{\delta}_{DR}$, we either misspecified $\pi(w_i;\eta)$ as a constant $\pi(w_i;\eta) = \text{expit}(\eta_0)$ for the IPW component or misspecified $g(\mathbf{w}_i,\gamma)$ as a constant $g(\mathbf{w}_i,\gamma) = \Phi(\gamma_0)$ for the MSI component. Let $\widehat{\delta}_{DR}^{IPW}$ ($\widehat{\delta}_{DR}^{MSI}$) denote the resulting DR estimator with the IPW (MSI) component specified correctly. Shown in Table 3 are the estimated $\vartheta$ for the FRM models for the DR estimator for the two scenarios, along with asymptotic and empirical standard errors for different sample sizes under 1,000 Monte Carlo samples. Both DR estimates were close to the true $\delta$ = 0.5 across all sample sizes. The empirical type I errors were close to the nominal level $\alpha$ = 0.05, albeit exhibiting a small upward bias for $n$ = 50. Table 3 also includes estimates from $\widehat{\delta}_{IPW}$ under misspecified propensity score and $\widehat{\delta}_{MSI}$ under misspecified outcome model for $n$ = 400 (large sample size to reduce sampling variability). As expected, both $\widehat{\delta}_{IPW}$ and $\widehat{\delta}_{MSI}$ showed a significant amount of bias. Table 4 summarizes the percentage of bias under different simulation scenarios. We also used the simulation data to compare power across the three FRM-based estimators with the traditional MWWRST ($\hat{\delta}_{MWW}$). We set $\beta_1$ = 1, while keeping the value for the rest parameters unchanged, and test the following hypothesis:

$$H_0 : Pr(y_{i1} \leq y_{j0}) = 0.5, \qquad vs. \qquad H_a : Pr(y_{i1} \leq y_{j0}) = a.$$

Under this setting, a is equal to 0.217 for the alternative $H_a$. Shown in Table 5 are power estimates for the three FRM-based causal estimators and the traditional MWWRST under the different sample sizes. As expected, power grew, as the sample size increased for all the estimators. Among the three causal estimators, $\hat{\delta}_{IPW}$ and $\hat{\delta}_{MSI}$ had comparable power, while $\hat{\delta}_{DR}$ had the largest power across all the sample sizes, which are consistent with the smallest standard errors observed earlier for this doubly robust estimator in Table 2. Power estimates for the traditional MWWRST were very different from those for the causal estimators, indicating that this noncausal estimator cannot be used to provide approximating power for any of the causal estimators.

### *5.2. Numerical Study*

We illustrate how the proposed approach may be used to address outliers using data from the Reach for Health (RFH) Study, a randomized control weight-loss trial to improve physical activities for non-diabetic breast cancer survivors conducted at the University of California San Diego. By using a randomized control trial to create confounders, we can see how the proposed approach addresses confounding effects in the presence of outliers in real studies without being confounded by hidden bias [6, 7].

The RFH study has four arms that included a total of 333 participants that were overweight/obsese (BMI $\geq 25 kg/m^2$) and diagnosed with stage 1, 2, or 3 breast cancer within the past 10 years. The four arms trial used a $2 \times 2$ factorial design with all participants randomly assigned to weight loss counseling vs. educational materials and to Metformin vs. placebo, with all interventions conducted over 6 months [22].

For illustration purposes, we combine the two medication groups (Metformin vs. placeb subjects) within each of the lifestyle intervention groups (weight loss counseling vs. educational materials) to consider only effects of the lifestyle intervention. For the subjects included in this analysis, their demographics are shown in Table 6. The weight loss counseling received lifestyle intervention consisted of 12 motivational interview calls from trained lifestyle coaches, while the educational materials group was given the 2010 US Dietary Guidelines. The general health score is derived from SF-36, which is a self-report questionnaire, with total scores ranging from 0 to 100 and with higher scores corresponding to better health.

Assessments of behavioral outcomes were recorded on a daily basis. The primary interest was to compare subjects' weighted summation of day-level Moderate and Vigorous Physical Activity (MVPA) count between the weight loss counseling and educational materials group during the 6-month intervention period. For each subject $i$, the weighted summation day-level MVPA count is calculated by $\sum_{ijk} I(X_{ijk} \geq 1952) \times X_{ijk}$, where $X_{ijk}$ denotes the activity count for the $k$th minute for subject $i$ at day $j$ and 1952 the threshold for counting as one minute of MVPA.

A common and challenging issue with the activity data is extreme values recorded by the device for subjects' activity levels. Shown in Table 1 are sample means and standard deviations of subjects' activity levels at the end of intervention, along with maximum (Max), interquartile range (IQR) and p-value from the two-sample t-test, to assess intervention effects for breast cancer survivors in a randomized control weightloss study. There are clearly outliers in both groups based on the common winsorizing approach, e.g., 3 times of IQR criterion [23]. While the t-test fails to capture any significant difference, the MWWRST shows a significant difference in mean rank between the two groups. The outliers in this study have a significant impact on study findings with important implications for clinical practice. The common approach of winsorizing outliers induces subjective opinions and rank-based methods such as the MWWRST objectively address this issue.

To use data from this RCT to illustrate the proposed approach for its ability to address confounding bias, we assumed a hypothetical scenario in which the study intervention was not efficacious and the observed group activity difference was the result of confounding bias in selecting subjects who were more (less) likely to exercise for the intervention (control) group. We then selected four covariates, age at diagnosis, BMI, high cholesterol (a binary with 0 indicating low and 1 indicating high cholesterol) and general health score (higher indicating better health), as such confounders. We assigned the lowest 166 values of age at diagnosis, BMI and high cholesterol, and the highest 166 values of general health score to the intervention group. The remaining 167 values of the four covariates were assigned to the control group.

Shown in Table 7 are the means (percent) and standard deviations of the three (one) continuous (binary) covariates. As expected, the intervention group was healthier compared to the control group with respect to the four covariates, and thus was more likely to exercise than the control group. Since we only changed values of the four covariates for the study subjects, the traditional MWWRST yielded the same test statistic and p-value as shown in Table 1. But, the significant difference of activity level between the groups now was the result of confounding bias due to the four covariates, rather than the intervention effect.

To address this bias, we applied the proposed doubly robust approach by modeling the effects of the confounders through the IPW and MSI components by modeling the propensity score and outcome regression using the respective semiparametric GLM and FRM with the logit link. The estimate of $\delta_{DR}$ was 0.501 with standard error = 0.043 and p-value = 0.973. Thus, the doubly robust estimator successfully addressed the biasing effects of the four confounders and indicated no treatment difference between the two groups of this hypothetical observational study.

## 6. Discussion

In this paper, we developed a doubly robust estimator to address the limitations of existing alternatives for more robust and reliable inferences when applying the MWWRST to observational study data. We investigated the performance of the proposed approach through both simulated and real data. The simulation study results demonstrated good performances even for samples as small as 50 when one of the propensity score and outcome regression model is correctly specified. The results from the real weight-loss trial showed that in addition to the doubly robust properties, the proposed estimator also effectively addressed outliers.

The proposed estimator is limited to cross-sectional study data. Work is currently underway to extend this approach to longitudinal studies with missing data to facilitate causal inferences for more complex study data arising in biomedical, clinical, epidemiological, and psychosocial research.

## Appendix A

We provide details of deriving the working variances in (15) for UGEE and a proof of Theorem 1.

### 1. Derivation of working variances in (15)

It is readily checked $V_{i1}$ and $V_{i2}$ in (15) are as below:

$$V_{i1} = Var(f_{i1} \mid \mathbf{w_i}) = \frac{1}{4}[\pi_i (1-\pi_i) + \pi_j (1-\pi_j)]$$

$$V_{i2} = Var(f_{i2} \mid \mathbf{w_i}) = \frac{1}{4}[g(\mathbf{w_i};\gamma)(1-g(\mathbf{w_i};\gamma)) + g(\mathbf{w_{i^c}};\gamma)(1-g(\mathbf{w_{i^c}};\gamma))].$$

To simplify the evaluation of $V_{i3}$ in (15), let:

$\pi_{\mathbf{i}} = \pi_i(1 - \pi_j), \pi_{\mathbf{j}} = \pi_j(1 - \pi_i)$

$r_{\mathbf{i}} = z_i(1 - z_j), r_{\mathbf{j}} = z_j(1 - z_i)$ Given

$$f_{\mathbf{i}3} = \frac{1}{2}[\frac{r_{\mathbf{i}}}{\pi_{\mathbf{i}}} I(y_{i1} \leq y_{j0}) + \left(1 - \frac{r_{\mathbf{i}}}{\pi_{\mathbf{i}}}\right) g(\mathbf{w}_{\mathbf{i}}, \boldsymbol{\gamma}) + \frac{r_{\mathbf{j}}}{\pi_{\mathbf{j}}} I(y_{j1} \leq y_{i0}) + \left(1 - \frac{r_{\mathbf{j}}}{\pi_{\mathbf{j}}}\right) g(\mathbf{w}_{\mathbf{i}^c}, \boldsymbol{\gamma})],$$

it follows from the iterated conditional expectation that:

$$\begin{aligned}
V_{\mathbf{i}3} &= Var(f_{\mathbf{i}3} \mid \mathbf{w}_{\mathbf{i}}) = E\left(f_{\mathbf{i}3}^2 \mid \mathbf{w}_{\mathbf{i}}\right) - E^2\left(f_{\mathbf{i}3} \mid \mathbf{w}_{\mathbf{i}}\right) \\
&= \frac{1}{4}\left\{E\left[\frac{r_{\mathbf{i}}^2}{\pi_{\mathbf{i}}^2}(I(y_{i1} \leq y_{j0}) - g(\mathbf{w}_{\mathbf{i}}, \gamma))^2 | \mathbf{w}_{\mathbf{i}}\right] + E\left[\frac{r_{\mathbf{j}}^2}{\pi_{\mathbf{j}}^2}(I(y_{j1} \leq y_{i0}) - g(\mathbf{w}_{\mathbf{i}^c}, \gamma))^2 | \mathbf{w}_{\mathbf{i}}\right]\right\} \\
&= \frac{1}{4}E\left[\frac{r_{\mathbf{i}}^2}{\pi_{\mathbf{i}}^2} g(\mathbf{w}_{\mathbf{i}}, \boldsymbol{\gamma})(1 - g(\mathbf{w}_{\mathbf{i}}, \boldsymbol{\gamma})) + \frac{r_{\mathbf{j}}^2}{\pi_{\mathbf{j}}^2} g(\mathbf{w}_{\mathbf{i}^c}, \boldsymbol{\gamma})(1 - g(\mathbf{w}_{\mathbf{i}^c}, \boldsymbol{\gamma})) | \mathbf{w}_{\mathbf{i}}\right] \\
&= \frac{1}{4}E\left[\frac{(z_i(1 - z_j))^2}{(\pi_i(1 - \pi_j))^2} g(\mathbf{w}_{\mathbf{i}}, \boldsymbol{\gamma})(1 - g(\mathbf{w}_{\mathbf{i}}, \boldsymbol{\gamma})) + \frac{(z_j(1 - z_i))^2}{(\pi_j(1 - \pi_i))^2} g(\mathbf{w}_{\mathbf{i}^c}, \boldsymbol{\gamma})(1 - g(\mathbf{w}_{\mathbf{i}^c}, \boldsymbol{\gamma})) | \mathbf{w}_{\mathbf{i}}\right] \\
&= \frac{1}{4}E\left[\frac{z_i(1 - z_j)}{(\pi_i(1 - \pi_j))^2} g(\mathbf{w}_{\mathbf{i}}, \boldsymbol{\gamma})(1 - g(\mathbf{w}_{\mathbf{i}}, \boldsymbol{\gamma})) + \frac{z_j(1 - z_i)}{(\pi_j(1 - \pi_i))^2} g(\mathbf{w}_{\mathbf{i}^c}, \boldsymbol{\gamma})(1 - g(\mathbf{w}_{\mathbf{i}^c}, \boldsymbol{\gamma})) | \mathbf{w}_{\mathbf{i}}\right] \\
&= \frac{1}{4}\left[\frac{1}{\pi_i(1 - \pi_j)} g(\mathbf{w}_{\mathbf{i}}, \boldsymbol{\gamma})(1 - g(\mathbf{w}_{\mathbf{i}}, \boldsymbol{\gamma})) + \frac{1}{\pi_j(1 - \pi_i)} g(\mathbf{w}_{\mathbf{i}^c}, \boldsymbol{\gamma})(1 - g(\mathbf{w}_{\mathbf{i}^c}, \boldsymbol{\gamma}))\right].
\end{aligned}$$

## 2. Proof of Theorem 1

Let $\mathbf{x}_i = \{y_i, z_i, \mathbf{w}_i\}$, $\mathbf{w}_\mathbf{i} = \{\mathbf{w}_{i_1}, \mathbf{w}_{i_2}\}$ and $\mathbf{x}_\mathbf{i} = \{\mathbf{x}_{i_1}, \mathbf{x}_{i_2}\}$. $(1 \le i \le n$, $\mathbf{i} = (i_1, i_2) \in C_2^n)$. We assume that $\boldsymbol{\theta} = \left(\boldsymbol{\eta}^\top, \boldsymbol{\gamma}^\top, \delta\right)^\top$ is a $q$–dimensional vector. Then $\mathbf{U}_n(\vartheta) = (U_{n,1}(\vartheta), \ldots, U_{n,q}(\vartheta))^\top$ is a $q$–dimensional random vector. Without loss of generality, we consider the normalized quantity $\binom{n}{2}^{-1} \sum_{\mathbf{i} \in C_2^n} \mathbf{U}_{n\mathbf{i}}(\boldsymbol{\theta})$ and continue to denote the normalized quantity as $\mathbf{U}_n(\vartheta)$ for notational brevity.

We first assume that $\alpha$ for the working correlation $R(\alpha)$ is known.

### 2.1. Consistency of $\widehat{\boldsymbol{\theta}}_n$

Consider a neighborhood $N(\vartheta)$ of $\vartheta$. We assume that $\mathbf{U}_n(\vartheta)$ has continuous second-order partial derivatives and there exists an integrable function $h(\mathbf{x}_\mathbf{i})$ with $Var(h(\mathbf{x}_\mathbf{i})) < \infty$ such that

$$\left|\frac{\partial^2}{\partial \lambda_k \partial \lambda_l} U_{n\mathbf{i},j}(\boldsymbol{\lambda})\right| \le h(\mathbf{x}_\mathbf{i}), \quad \text{for all} \quad \lambda \in N(\vartheta), \quad 0 \le k, l \le 2, \quad k+l = 2, \quad 1 \le j \le q, \tag{21}$$

where $U_{n\mathbf{i},j}$ denotes the $j$th component of $\mathbf{U}_{n\mathbf{i}}(\vartheta)$. Under (21), $Var(U_{n\mathbf{i},j}(\lambda)) < \infty$ for all $\lambda \in N(\vartheta)$, $1 \le j \le q$. Then UGEE estimators $\vartheta_\mathrm{b}$ are consistent and asymptotically normal under the above regularity conditions.

By the theory of multivariate U-statisitcs (Kowalski and Tu, 2007), $\mathbf{U}_n(\vartheta)$ is consistent and asymptotically normal, i.e.,

$$\mathbf{U}_n(\vartheta) \to_p \mathbf{0}, \qquad \sqrt{n}\mathbf{U}_n(\vartheta) \to_p N(\mathbf{0}, \boldsymbol{\Sigma}_U), \tag{22}$$

where $\boldsymbol{\Sigma}_U$ is the $q \times q$ asymptotic variance of $\mathbf{U}_n(\vartheta)$. Let

$$h_n = \binom{n}{2}^{-1} \sum_{\mathbf{i} \in C_2^n} h(\mathbf{x}_\mathbf{i}), \quad B(\boldsymbol{\theta}) = E\left(\frac{\partial \mathbf{U}_{n\mathbf{i}}(\boldsymbol{\theta})}{\partial \boldsymbol{\lambda}}\right).$$

Again, by the theory of multivariate U-statisitcs, we have:

$$h_n = E(h(\mathbf{x}_\mathbf{i})) + o_p(1), \quad \frac{\partial \mathbf{U}_n(\boldsymbol{\theta})}{\partial \boldsymbol{\lambda}} = B(\boldsymbol{\theta}) + \mathbf{o}_p(1), \tag{23}$$

where $o_p(1)$ denotes the stochastic scalar $o(1)$ and $\mathbf{o}_p(1)$ denotes the stochastic matrix $\mathbf{o}(1)$ (Kowalski and Tu, 2007).

By the mean-value theorem for vector-valued functions (Feng et al., 2014) and (23), we have:

$$U_{n,j}(\boldsymbol{\lambda}) = U_{n,j}(\boldsymbol{\theta}) + \frac{\partial}{\partial \boldsymbol{\lambda}} U_{n,j}(\boldsymbol{\theta})(\boldsymbol{\lambda} - \boldsymbol{\theta}) + R_{n,j}^{(2)}\left(\boldsymbol{\lambda} - \boldsymbol{\theta}, \boldsymbol{\xi}_j\right),$$

or in a vector form:

$$\begin{aligned}\mathbf{U}^n(\boldsymbol{\lambda}) &= \mathbf{U}_n(\boldsymbol{\theta}) + \frac{\partial}{\partial \boldsymbol{\lambda}} \mathbf{U}_n(\boldsymbol{\theta})(\boldsymbol{\lambda} - \boldsymbol{\theta}) + \mathbf{R}_n^{(2)}(\boldsymbol{\lambda} - \boldsymbol{\theta}, \boldsymbol{\xi}) \\ &= \mathbf{U}_n(\boldsymbol{\theta}) + B(\boldsymbol{\theta})(\boldsymbol{\lambda} - \boldsymbol{\theta}) + \mathbf{R}_n^{(2)}(\boldsymbol{\lambda} - \boldsymbol{\theta}, \boldsymbol{\xi}) + \mathbf{o}_p(1),\end{aligned} \tag{24}$$

where

$$\mathbf{R}_n^{(2)}(\boldsymbol{\lambda}-\boldsymbol{\theta},\boldsymbol{\xi}) = \left(R_{n,1}^{(2)}(\boldsymbol{\lambda}-\boldsymbol{\theta},\boldsymbol{\xi}_1),\ldots,R_{n,q}^{(2)}(\boldsymbol{\lambda}-\boldsymbol{\theta},\boldsymbol{\xi}_q)\right)^\top, \tag{25}$$

$$R_{n,j}^{(2)}(\boldsymbol{\lambda}-\boldsymbol{\theta},\boldsymbol{\xi}_j) = \frac{1}{2}\sum_{l=1}^{q}\sum_{k=1}^{q}\binom{2}{k,l}\frac{\partial^2}{\partial\lambda_k\partial\lambda_l}U_{n,j}(\boldsymbol{\xi}_j)(\lambda_k-\theta_k)(\lambda_l-\theta_l)$$

$$\boldsymbol{\xi} = \left\{\boldsymbol{\xi}_j; 1\le j\le q\right\},\quad \boldsymbol{\xi}_j\in N(\boldsymbol{\theta}),\quad 1\le j\le q.,$$

By (21) and (23), we can express $\mathbf{R}_n^{(2)}(\lambda - \vartheta,\xi)$ and $R_{n,j}^{(2)}(\boldsymbol{\lambda}-\boldsymbol{\theta},\boldsymbol{\xi}_j)$ as:

$$\begin{aligned} R_{n,j}^{(2)}(\boldsymbol{\lambda}-\boldsymbol{\theta},\boldsymbol{\zeta}_j) &= \frac{1}{2}\sum_{l=1}^{q}\sum_{k=1}^{q}\binom{2}{k,l}\zeta_{k,l,j}\left[E(h(\mathbf{x_i}))+o_p(1)\right](\lambda_k-\theta_k)(\lambda_l-\theta_l) \\ &= \frac{1}{2}\sum_{l=1}^{q}\sum_{k=1}^{q}\binom{2}{k,l}\zeta_{k,l,j}E(h(\mathbf{x_i}))(\lambda_k-\theta_k)(\lambda_l-\theta_l)+o_p(1) \\ &= H_j(\boldsymbol{\lambda}-\boldsymbol{\theta},\boldsymbol{\zeta}_j)+o_p(1), \\ \boldsymbol{\zeta} &= \left\{\boldsymbol{\zeta}_j;1\le j\le q\right\},\quad \boldsymbol{\zeta}_j=\{\zeta_{k,l,j};1\le k,l\le q\},\quad |\zeta_{k,l,j}|<1 \\ & 1\le k,l,j\le q, \end{aligned} \tag{26}$$

Further, for $1 \le j \le q$, we have:

$$\begin{aligned} \left|H_j((\boldsymbol{\lambda}-\boldsymbol{\theta}),\boldsymbol{\zeta}_j)\right| &\le \frac{1}{2}\sum_{l=1}^{q}\sum_{k=1}^{q}\binom{2}{k,l}|\zeta_{k,l,j}|\left|E(h(\mathbf{x_i}))(\lambda_k-\theta_k)(\lambda_l-\theta_l)\right| \\ &= \frac{1}{2}\sum_{l=1}^{q}\sum_{k=1}^{q}\binom{2}{k,l}\left|E(h(\mathbf{x_i}))\right|\left|(\lambda_k-\theta_k)(\lambda_l-\theta_l)\right|. \end{aligned} \tag{27}$$

Let

$$\mathbf{H}(\boldsymbol{\lambda}-\boldsymbol{\theta},\boldsymbol{\zeta}) = \left(H_1(\boldsymbol{\lambda}-\boldsymbol{\theta},\boldsymbol{\zeta}_1),\ldots,H_q(\boldsymbol{\lambda}-\boldsymbol{\theta},\boldsymbol{\zeta}_q)^\top\right).$$

Then we can express $\mathbf{R}_n^{(2)}(\lambda - \vartheta,\xi)$ as:

$$\mathbf{R}_n^{(2)}(\lambda - \vartheta,\zeta) = \mathbf{H}(\lambda - \vartheta,\zeta) + \mathbf{o}_p(1) \tag{28}$$

By (22), (24) and (28), we have:

$$\mathbf{U}_n(\lambda) = B(\vartheta)(\lambda - \vartheta) + \mathbf{H}(\lambda - \vartheta,\zeta) + \mathbf{o}_p(1). \tag{29}$$

By multiplying both sides by $F(\vartheta) = B^{-1}(\vartheta)$, we obtain from (29):

$$\begin{aligned} \mathbf{W}_n(\lambda) = F(\vartheta)\mathbf{U}_n(\lambda) &= \lambda - \vartheta + F(\vartheta)\mathbf{H}(\lambda - \vartheta,\zeta) + \mathbf{o}_p(1) \\ &= \lambda - \vartheta + F(\vartheta)\mathbf{H}(\lambda - \vartheta,\zeta) + \mathbf{z}_n \\ &= \lambda - \vartheta + \mathbf{G}(\lambda - \vartheta,\vartheta,\zeta) + \mathbf{z}_n \end{aligned} \tag{30}$$

where $\mathbf{z}_n = \mathbf{o}_p(1)$. Below we construct consistent estimators $\widehat{\boldsymbol{\theta}}_n$ of $\vartheta$ based on $\mathbf{W}_n(\lambda) = \mathbf{0}$, which implies that $\widehat{\boldsymbol{\theta}}_n$ is also a consistent estimator of $\vartheta$ based on

$\mathbf{U}_n(\lambda) = \mathbf{0}$.

Let

$$\boldsymbol{\lambda}_{1m} = \boldsymbol{\theta} - \frac{1}{m}, \quad \boldsymbol{\lambda}_{2m} = \boldsymbol{\theta} + \frac{1}{m}, \tag{31}$$

$$\Omega_{n,\frac{1}{m}} = \left\{\omega; \max |z_{nj}(\omega)| < \frac{1}{4m}, 1 \le j \le q\right\},$$

$$C = \max_{1\le i,j\le q} |F_{ij}(\boldsymbol{\theta})|, \quad m \ge 1,$$

where $\mathbf{1}_q$ denotes a $q \times 1$ column vector of 1's, $z_{nj}$ is the $j$th component of $\mathbf{z}_n$ and $F_{ij}(\vartheta)$ is the $ij$th element of $F(\vartheta)$. For $\lambda_{sm}$ ($s = 1,2$), by (27) we have for each component $G_j\left(\boldsymbol{\lambda}_{sm} - \boldsymbol{\theta}, \boldsymbol{\theta}, \boldsymbol{\zeta}_j\right)$ of $\mathbf{G}(\lambda_{sm}-\vartheta,\vartheta,\zeta)$:

$$\begin{aligned}\left|G_j\left(\boldsymbol{\lambda}_{sm} - \boldsymbol{\theta}, \boldsymbol{\theta}, \boldsymbol{\zeta}_j\right)\right| &\le C\sum_{j=1}^{q}\left|H_j\left(\left(\pm\frac{1}{m}\right)\mathbf{1}_q, \boldsymbol{\zeta}_j\right)\right| \\ &\le \frac{C}{2}\sum_{j=1}^{q}\sum_{l=1}^{q}\sum_{k=1}^{q}\binom{2}{k,l}\left|E\left(h\left(\mathbf{x_i}\right)\right)\right|\left(\frac{1}{m}\right)^2 \\ &= D\left(\frac{1}{m}\right)^2,\end{aligned}$$

where

$$D = \frac{C}{2}\sum_{j=1}^{q}\sum_{l=1}^{q}\sum_{k=1}^{q}\binom{2}{k,l}\left|E\left(h\left(\mathbf{x_i}\right)\right)\right|.$$

Thus for sufficiently large $m$ (e.g., $m > 4D$) we have:

$$\left|G_j\left(\boldsymbol{\lambda}_{sm}, \boldsymbol{\zeta}_j\right)\right| < \frac{1}{4m}, \quad s = 1,2, \quad 1 \le j \le q,$$
$$\boldsymbol{\lambda}_{sm} \in N(\boldsymbol{\theta}) \tag{32}$$

Since $\mathbf{z}_n = \mathbf{o}_p(1)$, for such $m$ we can find $N_{\frac{1}{m}}$ such that

$$\Pr\left[\Omega_{n,\frac{1}{m}}\right] \ge 1 - \frac{1}{m}, \qquad \text{for all } n \ge N_{\frac{1}{m}}. \tag{33}$$

For $\omega \in \Omega_{n,\frac{1}{m}}$, it follows from (30), (31) and (32) that for all $1 \le j \le q$

$$w_{nj}(\boldsymbol{\lambda}_{1m}) = -\frac{1}{m} + G_j\left(\boldsymbol{\lambda}_{1m}, \boldsymbol{\zeta}_j\right) + z_{nj} < -\frac{1}{m} + \frac{1}{4m} + \frac{1}{4m} = -\frac{1}{2m} < 0$$

$$w_{nj}(\boldsymbol{\lambda}_{2m}) = \frac{1}{m} + G_j\left(\boldsymbol{\lambda}_{2m}, \boldsymbol{\zeta}_j\right) + z_{nj} > \frac{1}{m} - \frac{1}{4m} - \frac{1}{4m} = \frac{1}{2m} > 0.$$

Since $\mathbf{w}_n(\lambda)$ is continuous in $\lambda$, there exists $\eta_m \in N(\vartheta)$ such that $\mathbf{w}_n(\eta_m) = \mathbf{0}$. For $n > N_{\frac{1}{m}}$, let $\widehat{\vartheta}_n = \inf\{\|\eta_m\| : \{\eta_m \in N(\vartheta)\}^{\top}\{\mathbf{w}_n(\eta_m) = \mathbf{0}\}\}.$

Then, $\widehat{\boldsymbol{\theta}}_n$ is a sequence of random vectors. Further, by (33), we have:

$$\Pr\left[\left\|\widehat{\boldsymbol{\theta}}_n - \boldsymbol{\theta}\right\| \geq \frac{1}{m}\right] \leq \Pr\left[\Omega^c_{n,\frac{1}{m}} \text{ for } n \geq N_{\frac{1}{m}}\right] \leq \frac{1}{4m} \to 0 . \tag{34}$$

Thus $\widehat{\boldsymbol{\theta}}_n$ is a consistent estimate of $\vartheta$.

We now consider the case where $\alpha$, a $p$−dimensional vector, is unknown and is estimated by a $\sqrt{n}$−consistent estimator, $\widehat{\alpha}$, i.e., $\sqrt{n}(\widehat{\alpha} - \alpha) = \mathbf{O}_p(1)$, where $\mathbf{O}_p(1)$ denotes a vector of the same dimension as $\alpha$ with all its components stochastically bounded (Kowalski and Tu, 2007). We consider the augmented vector $\left(\boldsymbol{\theta}^{\top}, \boldsymbol{\alpha}^{\top}\right)^{\top}$ so $\boldsymbol{\lambda}$ has the dimension of $q + p$. In this case, we partition $\boldsymbol{\lambda} = \left(\boldsymbol{\lambda}_1^{\top}, \boldsymbol{\lambda}_2^{\top}\right)^{\top}$, where $\lambda_1$ is the $q$−dimensional vector corresponding to $\vartheta$ and $\lambda_2$ $p$−dimensional vector corresponding $\alpha$. We assume that the regularity conditions in (21) holds true with respect to the augmented $\lambda$.

Similar to (24) - (29) for the case of known $\alpha$, we have:

$$\begin{aligned}\mathbf{U}_n(\boldsymbol{\lambda}) &= \mathbf{U}_n(\boldsymbol{\theta},\boldsymbol{\alpha}) + \frac{\partial}{\partial\boldsymbol{\lambda}_1}\mathbf{U}_n(\boldsymbol{\theta},\boldsymbol{\alpha})(\boldsymbol{\lambda}_1 - \boldsymbol{\theta}) + \frac{\partial}{\partial\boldsymbol{\lambda}_2}\mathbf{U}_n(\boldsymbol{\theta},\boldsymbol{\alpha})(\boldsymbol{\lambda}_2 - \boldsymbol{\alpha}) + \mathbf{R}_n^{(2)}(\boldsymbol{\lambda}_1 - \boldsymbol{\theta}, \boldsymbol{\lambda}_2 - \boldsymbol{\alpha}, \boldsymbol{\zeta}) \\ &= B(\boldsymbol{\theta},\boldsymbol{\alpha})(\boldsymbol{\lambda}_1 - \boldsymbol{\theta}) + \frac{\partial}{\partial\boldsymbol{\lambda}_2}\mathbf{U}_n(\boldsymbol{\theta},\boldsymbol{\alpha})(\boldsymbol{\lambda}_2 - \boldsymbol{\alpha}) + \mathbf{R}_n^{(2)}(\boldsymbol{\lambda}_1 - \boldsymbol{\theta}, \boldsymbol{\lambda}_2 - \boldsymbol{\alpha}, \boldsymbol{\zeta}) + \mathbf{o}_p(1),\end{aligned} \tag{35}$$

where $B(\vartheta,\alpha)$ and $\mathbf{R}_n^{(2)}(\lambda - \vartheta, \xi)$ in this case is given by:

$$\begin{aligned} B(\boldsymbol{\theta},\boldsymbol{\alpha}) &= E\left(\frac{\partial \mathbf{U}_{n\mathbf{i}}(\boldsymbol{\theta})}{\partial\boldsymbol{\lambda}_1}\right) \\ \mathbf{R}_n^{(2)}(\boldsymbol{\lambda}_1 - \boldsymbol{\theta}, \boldsymbol{\lambda}_2 - \boldsymbol{\alpha}, \boldsymbol{\zeta}) &= \mathbf{R}_{1n}^{(2)}(\boldsymbol{\lambda}_1 - \boldsymbol{\theta}, \boldsymbol{\zeta}) + \mathbf{R}_{2n}^{(2)}(\boldsymbol{\lambda}_1 - \boldsymbol{\theta}, \boldsymbol{\lambda}_2 - \boldsymbol{\alpha}, \boldsymbol{\zeta}) + \mathbf{R}_{3n}^{(2)}(\boldsymbol{\lambda}_2 - \boldsymbol{\alpha}, \boldsymbol{\zeta}) \\ R_{1n,j}\left(\boldsymbol{\lambda}_1 - \boldsymbol{\theta}, \boldsymbol{\zeta}_j\right) &= \frac{1}{2}\sum_{l=1}^{q}\sum_{k=1}^{q}\binom{2}{k,l}\zeta_{k,l,j}\left[E(h(\mathbf{x_i})) + o_p(1)\right](\lambda_{1k} - \theta_k)(\lambda_{1l} - \theta_l) \\ R_{2n,j}\left(\boldsymbol{\lambda}_1 - \boldsymbol{\theta}, \boldsymbol{\lambda}_2 - \boldsymbol{\alpha}, \boldsymbol{\zeta}_j\right) &= \sum_{l'=1}^{p}\sum_{k=1}^{q}\binom{2}{k,l'}\zeta_{k,l',j}\left[E(h(\mathbf{x_i})) + o_p(1)\right](\lambda_{1k} - \theta_k)(\lambda_{2l'} - \alpha_{l'}) \\ R_{3n,j}\left(\boldsymbol{\lambda} - \boldsymbol{\alpha}, \boldsymbol{\zeta}_j\right) &= \frac{1}{2}\sum_{l'=1}^{p}\sum_{k'=1}^{p}\binom{2}{k',l'}\zeta_{k',l',j}\left[E(h(\mathbf{x_i})) + o_p(1)\right](\lambda_{2k'} - \alpha_{k'})(\lambda_{2l'} - \alpha_{l'}) .\end{aligned} \tag{36}$$

By substituting $\widehat{\alpha}$ in place of $\alpha$, we have:

$$R_{2n,j}\left(\boldsymbol{\lambda}_1 - \boldsymbol{\theta}, \boldsymbol{\lambda}_2 - \boldsymbol{\alpha}, \boldsymbol{\zeta}_j\right) = \sum_{l=1}^{q} o_p(1)(\lambda_l - \theta_l), \quad R_{3n,j}\left(\boldsymbol{\lambda} - \boldsymbol{\alpha}, \boldsymbol{\zeta}_j\right) = o_p(1). \tag{37}$$

Thus by (36) and (37), we have:

$$R_{n,j}^{(2)}(\boldsymbol{\lambda}_1-\boldsymbol{\theta},\boldsymbol{\lambda}_2-\boldsymbol{\alpha},\boldsymbol{\xi}) = R_{1n,j}(\boldsymbol{\lambda}_1-\boldsymbol{\theta},\boldsymbol{\zeta}_j)+o_p(1)$$
$$= H_j(\boldsymbol{\lambda}_1-\boldsymbol{\theta},\boldsymbol{\zeta}_j)+o_p(1),$$

where $H_j(\boldsymbol{\lambda}_1-\boldsymbol{\theta},\boldsymbol{\zeta}_j)$ has the same form as the $H_j(\boldsymbol{\lambda}-\boldsymbol{\theta},\boldsymbol{\zeta}_j)$ in (26), except for the notational change from $\lambda$ to $\lambda_1$. By expressing the above in a vector form, we have:

$$\mathbf{R}_n^{(2)}(\boldsymbol{\lambda}_1-\boldsymbol{\theta},\boldsymbol{\lambda}_2-\boldsymbol{\alpha},\boldsymbol{\xi}) = \mathbf{H}(\boldsymbol{\lambda}_1-\boldsymbol{\theta},\boldsymbol{\zeta})+\mathbf{o}_p(1), \tag{38}$$

Let $s$ denote the dimension of the functional response $\mathbf{f_i}$ ( $s$ = 3 for the specific FRM in (14)). By the definition of $\mathbf{U}_{n\mathbf{i}}(\vartheta,\alpha)$, we have:

$$\frac{\partial}{\partial\boldsymbol{\lambda}_2}\mathbf{U}_n(\boldsymbol{\theta},\boldsymbol{\alpha}) = \sum_{\mathbf{i}\in C_2^n}\frac{\partial}{\partial\boldsymbol{\lambda}_2}\mathbf{U}_{n\mathbf{i}}(\boldsymbol{\theta},\boldsymbol{\alpha}) = \sum_{\mathbf{i}\in C_2^n}\frac{\partial}{\partial\boldsymbol{\lambda}_2}\left(D_{\mathbf{i}}V_{\mathbf{i}}^{-1}S_{\mathbf{i}}\right).$$

Let $A_{\mathbf{i}} = D_{\mathbf{i}}V_{\mathbf{i}}^{-1} = [a_{\mathbf{i},kl}]$, a $(q + p) \times s$ matrix with the element $a_{\mathbf{i},kl}m_{kl,j}$ in the $k$th row and $j$th column of $A_{\mathbf{i}}$. Since only $D_{\mathbf{i}}V_{\mathbf{i}}^{-1}$ involves $\alpha$, we have:

$$\frac{\partial}{\partial\boldsymbol{\lambda}_2}\left(D_{\mathbf{i}}V_{\mathbf{i}}^{-1}S_{\mathbf{i}}\right) = \left[\frac{\partial}{\partial\lambda_{2r}}\sum_{l=1}^{s}a_{\mathbf{i},kl}S_{\mathbf{i},l}\right] = \left[\sum_{l=1}^{s}\left(\frac{\partial}{\partial\lambda_{2r}}a_{\mathbf{i},kl}\right)S_{\mathbf{i},l}\right],$$

where $S_{\mathbf{i},l}$ is the $l$th component of $S_{\mathbf{i}}$ and $\sum_{l=1}^{s}\left(\frac{\partial}{\partial\lambda_{2r}}a_{\mathbf{i},kl}\right)S_{\mathbf{i},l}$ is the element in the $r$th row and $k$th column of $\frac{\partial}{\partial\boldsymbol{\lambda}_2}\left(D_{\mathbf{i}}V_{\mathbf{i}}^{-1}S_{\mathbf{i}}\right)$. By the iterated conditional expectation, we have:

$$E\left[\left(\frac{\partial}{\partial\lambda_{2r}}a_{\mathbf{i},kl}\right)S_{\mathbf{i},l}\right] = E\left\{E\left[\left(\frac{\partial}{\partial\lambda_{2r}}a_{\mathbf{i},kl}\right)S_{\mathbf{i},l}\mid\mathbf{w_i}\right]\right\}$$
$$= E\left[\left(\frac{\partial}{\partial\lambda_{2r}}a_{\mathbf{i},kl}\right)E\left(S_{\mathbf{i},l}\mid\mathbf{w_i}\right)\right]$$
$$= \mathbf{0}.$$

By the theory of multivariate U-statisitcs,

$$\frac{\partial}{\partial\boldsymbol{\lambda}_2}\mathbf{U}_n(\boldsymbol{\theta},\boldsymbol{\alpha}) = \mathbf{o}_p(1). \tag{39}$$

It follows from (35), (38) and (39) that

$$\mathbf{U}_n(\lambda) = B(\vartheta,\alpha)(\lambda_1-\vartheta)+\mathbf{H}(\lambda_1-\vartheta,\zeta)+\mathbf{o}_p(1). \tag{40}$$

The term $\mathbf{H}(\lambda_1-\vartheta,\zeta)$ has the same properties as $\mathbf{H}(\lambda-\vartheta,\zeta)$ in (29). Thus the consistency of $\widehat{\boldsymbol{\theta}}_n$ follows from the same argument through the steps in (30) - (34) for the case of known $\alpha$.

**2.2 Asymptotic Normality of $\widehat{\theta}_n$** From (24), we have:

$$\mathbf{U}_n(\boldsymbol{\lambda}) = \mathbf{U}_n(\boldsymbol{\theta}) + B(\boldsymbol{\theta})(\boldsymbol{\lambda} - \boldsymbol{\theta}) + \mathbf{R}_n^{(2)}(\boldsymbol{\lambda} - \boldsymbol{\theta}, \boldsymbol{\zeta}) + \mathbf{o}_p(1). \tag{41}$$

Let $M_j(\boldsymbol{\lambda} - \boldsymbol{\theta}, \boldsymbol{\zeta}_j) = [m_{kl,j}(\boldsymbol{\lambda} - \boldsymbol{\theta}, \boldsymbol{\zeta}_j)]$ be a $q \times q$ matrix with the element $m_{klj}$ in the $k$th row and $j$th column given by:

$$m_{kl,j}(\boldsymbol{\lambda} - \boldsymbol{\theta}, \boldsymbol{\zeta}_j) = \frac{1}{2}\binom{2}{k,l}\zeta_{k,l,j}E(h(\mathbf{x_i})), \quad 1 \leq, j \leq q. \tag{42}$$

From (26) and (42), we have:

$$\begin{aligned} R_{n,j}^{(2)}(\boldsymbol{\lambda} - \boldsymbol{\theta}, \boldsymbol{\zeta}_j) &= \frac{1}{2}\sum_{k=1}^{q}\sum_{l=1}^{q}\binom{2}{k,l}\zeta_{k,l,j}[E(h(\mathbf{x_i})) + o_p(1)](\lambda_k - \theta_k)(\lambda_l - \theta_l) \quad (43) \\ &= (\boldsymbol{\lambda} - \boldsymbol{\theta})^\top (M_j(\boldsymbol{\lambda} - \boldsymbol{\theta}, \boldsymbol{\zeta}_j) + \mathbf{o}_p(1))(\boldsymbol{\lambda} - \boldsymbol{\theta}), \end{aligned}$$

where $\mathbf{o}_p(1)$ is $q \times q$ matrix of $o_p(1)$. By substituting $\widehat{\boldsymbol{\theta}}_n$ in place of $\lambda$ and consistency of $\vartheta b_n$, we have:

$$\begin{aligned} R_{n,j}^{(2)}\left(\widehat{\boldsymbol{\theta}}_n - \boldsymbol{\theta}, \boldsymbol{\zeta}_j\right) &= \left(\widehat{\boldsymbol{\theta}}_n - \boldsymbol{\theta}\right)^\top \left(M_j\left(\widehat{\boldsymbol{\theta}}_n - \boldsymbol{\theta}, \boldsymbol{\zeta}_j\right) + \mathbf{o}_p(1)\right)\left(\widehat{\boldsymbol{\theta}}_n - \boldsymbol{\theta}\right) \\ &= \mathbf{o}_p(1)\left(M_j\left(\widehat{\boldsymbol{\theta}}_n - \boldsymbol{\theta}, \boldsymbol{\zeta}_j\right) + \mathbf{o}_p(1)\right)\left(\widehat{\boldsymbol{\theta}}_n - \boldsymbol{\theta}\right) \\ &= \mathbf{o}_p(1)\left(\widehat{\boldsymbol{\theta}}_n - \boldsymbol{\theta}\right) \end{aligned} \tag{44}$$

By substituting $\widehat{\boldsymbol{\theta}}_n$ in place of $\lambda$ and the fact $\mathbf{U}^n\left(\widehat{\boldsymbol{\theta}}_n\right) = \mathbf{0}$, it follows from (41) and (44) that

$$\begin{aligned} \mathbf{0} &= \mathbf{U}_n(\boldsymbol{\theta}) + B(\boldsymbol{\theta})\left(\widehat{\boldsymbol{\theta}}_n - \boldsymbol{\theta}\right) + \mathbf{o}_p(1)\left(\widehat{\boldsymbol{\theta}}_n - \boldsymbol{\theta}\right) \\ &= \mathbf{U}_n(\boldsymbol{\theta}) + [B(\boldsymbol{\theta}) + \mathbf{o}_p(1)]\left(\widehat{\boldsymbol{\theta}}_n - \boldsymbol{\theta}\right). \end{aligned} \tag{45}$$

Multiplying both sides of (44) by $\sqrt{n}$ and solving the resulting equations for $\sqrt{n}(\lambda - \vartheta)$ yields:

$$\sqrt{n}\left(\widehat{\boldsymbol{\theta}}_n - \boldsymbol{\theta}\right) = B^{-1}(\boldsymbol{\theta})\sqrt{n}\mathbf{U}_n(\boldsymbol{\theta}) + \mathbf{o}_p(1). \tag{46}$$

Since $\sqrt{n}\mathbf{U}_n(\vartheta)$ has the same asymptotic distribution as its projection:

$$\sqrt{n}\widehat{\mathbf{U}}_n(\boldsymbol{\theta}) = \frac{\sqrt{n}}{n}\sum_{i=1}^{n}\widehat{\mathbf{U}}_{n,i}(\boldsymbol{\theta}) = \frac{\sqrt{n}}{n}\sum_{i=1}^{n}2E(\mathbf{U}_{n,\mathbf{i}} \mid y_{i1}, y_{i0}, z_i, \mathbf{w}_i), \tag{47}$$

we can rewrite (45) as:

$$\sqrt{n}\left(\widehat{\boldsymbol{\theta}}_n - \boldsymbol{\theta}\right) = -[B(\boldsymbol{\theta})]^{-1}\sqrt{n}\widehat{\mathbf{U}}_n(\boldsymbol{\theta}) + \mathbf{o}_p(1).$$

√_

As $\sqrt{n}\widehat{\mathbf{U}}_n(\vartheta)$ has an asymptotic normal $N(\mathbf{0}, \Sigma_\theta(\vartheta) = 4\Sigma(\vartheta))$, it follows from Slutsky's theorem that

$$\sqrt{n}\left(\widehat{\boldsymbol{\theta}}_n - \boldsymbol{\theta}\right) \to_d N\left(\mathbf{0}, \Sigma_\theta(\boldsymbol{\theta}) = 4B^{-1}(\boldsymbol{\theta})\,\Sigma(\boldsymbol{\theta})\,B^{-1}(\boldsymbol{\theta})\right). \tag{48}$$

For the case of unknown $\alpha$ with a $\sqrt{n}$–consistent estimator $\widehat{\alpha}$, we again consider $\boldsymbol{\lambda} = \left(\boldsymbol{\theta}^\top, \boldsymbol{\alpha}^\top\right)^\top$ so $\boldsymbol{\lambda}$ the augmented vector has the dimension of $q + p$ and is partitioned into $\boldsymbol{\lambda} = \left(\boldsymbol{\lambda}_1^\top, \boldsymbol{\lambda}_2^\top\right)^\top$, where $\lambda_1$ is the $q$–dimensional vector corresponding to $\vartheta$ and $\lambda_2$ $p$–dimensional vector corresponding $\alpha$.

By applying the argument in (42) - (44) to $\mathbf{R}^{(2)}_{1n}(\boldsymbol{\lambda}_1 - \boldsymbol{\theta}, \boldsymbol{\zeta})$, $\mathbf{R}^{(2)}_{2n}(\boldsymbol{\lambda}_1 - \boldsymbol{\theta}, \boldsymbol{\lambda}_2 - \boldsymbol{\alpha}, \boldsymbol{\zeta})$ and $\mathbf{R}^{(2)}_{3n}(\boldsymbol{\lambda}_2 - \boldsymbol{\alpha}, \boldsymbol{\zeta})$ in (36) and the fact both $\widehat{\vartheta}_n$ and $\widehat{\alpha}$ are consistent, we obtain:

$$\begin{aligned}\mathbf{R}^{(2)}_{1n}\left(\widehat{\boldsymbol{\theta}}_n - \boldsymbol{\theta}, \boldsymbol{\zeta}\right) &= \mathbf{o}_p(1)\left(\widehat{\boldsymbol{\theta}}_n - \boldsymbol{\theta}\right)\\ \mathbf{R}^{(2)}_{2n}\left(\boldsymbol{\lambda}_1 - \boldsymbol{\theta}, \widehat{\boldsymbol{\alpha}} - \boldsymbol{\alpha}, \boldsymbol{\zeta}\right) &= \mathbf{o}_p(1)\left(\widehat{\boldsymbol{\theta}}_n - \boldsymbol{\theta}\right)\\ \mathbf{R}^{(2)}_{3n}\left(\widehat{\boldsymbol{\alpha}} - \boldsymbol{\alpha}, \boldsymbol{\zeta}\right) &= \mathbf{o}_p(1)\left(\widehat{\boldsymbol{\alpha}} - \boldsymbol{\alpha}\right).\end{aligned} \tag{49}$$

By substituting $\widehat{\boldsymbol{\theta}}_n$ in place of $\lambda_1$ and $\widehat{\alpha}$ in place of $\lambda_2$, we obtain from (35), (39) and (49):

$$\begin{aligned}\mathbf{U}_n\left(\widehat{\boldsymbol{\theta}}_n, \widehat{\boldsymbol{\alpha}}\right) &= \mathbf{U}_n(\boldsymbol{\theta}, \boldsymbol{\alpha}) + B(\boldsymbol{\theta})\left(\widehat{\boldsymbol{\theta}}_n - \boldsymbol{\theta}\right) + \mathbf{o}_p(1)\left(\widehat{\boldsymbol{\alpha}} - \boldsymbol{\alpha}\right) +\\ &+ \mathbf{o}_p(1)\left(\widehat{\boldsymbol{\theta}}_n - \boldsymbol{\theta}\right) + \mathbf{o}_p(1)\left(\widehat{\boldsymbol{\theta}}_n - \boldsymbol{\theta}\right) + \mathbf{o}_p(1)\left(\widehat{\boldsymbol{\alpha}} - \boldsymbol{\alpha}\right)\end{aligned} \tag{50}$$

Multiplying both sides of (50) by $\sqrt{n}$ and substituting $\mathbf{0}$ in place of $\mathbf{U}_n\left(\widehat{\boldsymbol{\theta}}_n, \widehat{\boldsymbol{\alpha}}\right)$ yields:

$$\begin{aligned}\mathbf{0} &= \sqrt{n}\mathbf{U}_n(\boldsymbol{\theta}, \boldsymbol{\alpha}) + B(\boldsymbol{\theta})\sqrt{n}\left(\widehat{\boldsymbol{\theta}}_n - \boldsymbol{\theta}\right) + \mathbf{o}_p(1)\sqrt{n}\left(\widehat{\boldsymbol{\alpha}} - \boldsymbol{\alpha}\right) +\\ &+ \mathbf{o}_p(1)\sqrt{n}\left(\widehat{\boldsymbol{\theta}}_n - \boldsymbol{\theta}\right) + \mathbf{o}_p(1)\sqrt{n}\left(\widehat{\boldsymbol{\theta}}_n - \boldsymbol{\theta}\right) + \mathbf{o}_p(1)\sqrt{n}\left(\widehat{\boldsymbol{\alpha}} - \boldsymbol{\alpha}\right).\end{aligned} \tag{51}$$

Since $\sqrt{n}(\widehat{\alpha} - \alpha) = \mathbf{O}_p(1)$ and $\sqrt{n}(\widehat{\alpha} - \alpha) = \mathbf{O}_p(1)$, (51) simplifies to:

$$\mathbf{0} = \sqrt{n}\mathbf{U}_n(\boldsymbol{\theta}, \boldsymbol{\alpha}) + B(\boldsymbol{\theta})\sqrt{n}\left(\widehat{\boldsymbol{\theta}}_n - \boldsymbol{\theta}\right) + \mathbf{o}_p(1) \tag{52}$$

Solving the equations in (52) for $\sqrt{n}\left(\widehat{\boldsymbol{\theta}}_n - \boldsymbol{\theta}\right)$ yields:

$$\sqrt{n}\left(\widehat{\boldsymbol{\theta}}_n - \boldsymbol{\theta}\right) = B^{-1}(\boldsymbol{\theta})\sqrt{n}\mathbf{U}_n(\boldsymbol{\theta}, \boldsymbol{\alpha}) + \mathbf{o}_p(1) \tag{53}$$

The conclusion follows by replacing $n\mathbf{U}_n(\boldsymbol{\vartheta},\boldsymbol{\alpha})$ with its projection) $\sqrt{n}\widehat{\mathbf{U}}_n(\boldsymbol{\theta},\boldsymbol{\alpha}$ and the asymptotic normality of the projection $\sqrt{n}\widehat{\mathbf{U}}_n(\boldsymbol{\theta},\boldsymbol{\alpha})$, as in (47)-(48) for the case of known $\boldsymbol{\alpha}$.

**Table 1.** Two sample t-test vs. MWWRST for group difference of Weighted Sum Activity Count of MVPA.

| Two sample t-test | | | |
|---|---|---|---|
| | Intervention mean(SD) | Control mean(SD) | p-value(t) |
| AC MVPA | 3595378 (1652517) | 3231656 (1634587) | 0.091 |
| Max/IQR | 5.53 | 5.80 | |
| **MWWRST** | | | |
| | Estimate(SE) | | p-value |
| $\delta$ | 0.430(0.035) | | 0.039 |

**Table 2.** Comparison of estimates of $\delta$ and standard errors (asymptotic vs. empirical) between FRM and traditional MWW approaches for the Simulation Study. For the Doubly Robust estimator, both IPW and MSI are correctly specified.

| | $\eta_0$ | $\eta_1$ | $\gamma_0$ | $\gamma_{11}$ | $\gamma_{10}$ | $\delta$ | $\alpha$ |
|---|---|---|---|---|---|---|---|
| | | | | $n = 50$ | | | |
| $\widehat{\delta}_{DR}$ | 1.087 (0.678/0.746) | -1.071 (0.613/0.678) | 0.093 (0.964/1.915) | -0.583 (0.825/1.285) | 0.590 (1.127/1.943) | 0.493 (0.085/0.080) | 0.059 |
| $\widehat{\delta}_{IPW}$ | 1.079 (0.679/0.737) | -1.068 (0.612/0.659) | | | | 0.513 (0.086/0.083) | 0.055 |
| $\widehat{\delta}_{MSI}$ | | | 0.101 (0.933/1.884) | -0.582 (0.792/1.117) | 0.601 (1.084/1.896) | 0.511 (0.084/0.083) | 0.063 |
| $\widehat{\delta}_{MWW}$ | | | | | | 0.562 (0.080/0.110) | 0.310 |
| | | | | $n = 200$ | | | |
| $\widehat{\delta}_{DR}$ | 1.010 (0.327/0.334) | -1.009 (0.304/0.314) | 0.045 (0.793/0.857) | -0.495 (0.596/0.624) | 0.511 (0.741/0.830) | 0.497 (0.040/0.040) | 0.049 |
| $\widehat{\delta}_{IPW}$ | 1.009 (0.328/0.335) | -1.006 (0.306/0.320) | | | | 0.502 (0.041/0.042) | 0.048 |
| $\widehat{\delta}_{MSI}$ | | | 0.046 (0.804/0.862) | -0.495 (0.607/0.648) | 0.508 (0.744/0.842) | 0.501 (0.041/0.041) | 0.051 |
| $\widehat{\delta}_{MWW}$ | | | | | | 0.558 (0.041/0.040) | 0.190 |
| | | | | $n = 400$ | | | |
| $\widehat{\delta}_{DR}$ | 1.004 (0.238/0.248) | -1.007 (0.215/0.223) | 0.041 (0.572/0.610) | -0.496 (0.485/0.505) | 0.505 (0.559/0.564) | 0.499 (0.037/0.037) | 0.050 |
| $\widehat{\delta}_{IPW}$ | 1.003 (0.238/0.250) | -1.007 (0.214/0.221) | | | | 0.501 (0.039/0.039) | 0.050 |
| $\widehat{\delta}_{MSI}$ | | | 0.039 (0.572/0.611) | -0.501 (0.488/0.512) | 0.504 (0.560/0.565) | 0.497 (0.038/0.039) | 0.052 |
| $\widehat{\delta}_{MWW}$ | | | | | | 0.559 (0.037/0.038) | 0.160 |

**Table 3.** Comparison estimates and standard errors (asymptotic vs. empirical) of doubly robust estimator with only one component correctly specified and IPW (MSI) estimator with incorrectly specified propensity (imputation) function.

| | $\eta_0$ | $\eta_1$ | $\gamma_0$ | $\gamma_{11}$ | $\gamma_{10}$ | $\delta$ | $\alpha$ |
|---|---|---|---|---|---|---|---|
| | | | | $n = 50$ | | | |
| $\widehat{\delta}_{DR}^{IPW}$ | 1.088 (0.698/0.787) | -1.112 (0.675/0.783) | 3.148 (0.975/1.963) | - | - | 0.508 (0.091/0.089) | 0.059 |
| $\widehat{\delta}_{DR}^{MSI}$ | -0.002 (0.286/0.295) | - | 0.106 (0.981/1.924) | -0.580 (0.862/1.227) | 0.573 (1.264/2.003) | 0.509 (0.088/0.087) | 0.064 |
| | | | | $n = 200$ | | | |
| $\widehat{\delta}_{DR}^{IPW}$ | 1.009 (0.336/0.349) | -1.005 (0.322/0.339) | 2.817 (0.815/1.052) | - | - | 0.502 (0.043/0.044) | 0.049 |
| $\widehat{\delta}_{DR}^{MSI}$ | -0.005 (0.138/0.142) | - | 0.047 (0.822/0.893) | -0.489 (0.646/0.703) | 0.507 (0.881/0.972) | 0.501 (0.042/0.043) | 0.051 |
| | | | | $n = 400$ | | | |
| $\widehat{\delta}_{DR}^{IPW}$ | 1.003 (0.241/0.253) | -1.006 (0.215/0.227) | 2.309 (0.751/0.803) | - | - | 0.501 (0.040/0.041) | 0.050 |
| $\widehat{\delta}_{DR}^{MSI}$ | -0.004 (0.100/0.100) | - | 0.038 (0.575/0.614) | -0.501 (0.487/0.513) | 0.505 (0.563/0.567) | 0.497 (0.041/0.041) | 0.051 |
| | | | 2.311 (0.763/0.842) | | | 0.563 (0.041/0.043) | 0.164 |
| $\widehat{\delta}_{IPW}$ | -0.003 (0.107/0.114) | - | - | - | - | 0.559 (0.041/0.042) | 0.131 |

$\delta b_{MSI}$

- - - -

**Table 4.** Comparison of percentage of bias between FRM and traditional MWW approaches for the Simulation Study.

| | $\hat{\delta}_{DR}$ | $\hat{\delta}_{IPW}$ | $\hat{\delta}_{MSI}$ | $\hat{\delta}_{DRIPW}$ | $\hat{\delta}_{DRMSI}$ | $\hat{\delta}_{MWW}$ |
|---|---|---|---|---|---|---|
| percentage of bias (n=50) | -1.4% | 2.6% | 2.2% | 1.6% | 1.8% | 12.4% |
| percentage of bias (n=200) | -0.6% | 0.4% | 0.2% | 0.4% | 0.2% | 11.6% |
| percentage of bias (n=400) | -0.2% | 0.2% | -0.6% | 0.2% | -0.6% | 11.8% |

**Table 5.** Power comparison between FRM-based double robust estimator and traditional MWW

| | n = 50 | n = 200 | n = 400 |
|---|---|---|---|
| $\delta_{IPW}$ | | 0.937 | 0.945 |
| $\hat{\delta}_{MSI}$ | 0.609 | 0.936 | 0.943 |
| $\hat{\delta}_{DR}$ | 0.618 | 0.939 | 0.948 |
| $\hat{\delta}_{MWW}$ | 0.725 | 0.853 | 0.891 |

**Table 6.** Demographics characteristics by groups.

| | Intervention (N=166) mean (SD) | Control (N=167) mean (SD) |
|---|---|---|
| Age at Enrollment | 62.7(7.07) | 62.5(7.01) |
| Age at Diagnosis | 60.1(7.08) | 60.0(6.80) |
| Year from Diagnosis to study enrollment | 2.7(2.04) | 2.5(1.79) |
| General Health (SF-36[a]) | 73.6(18.54) | 70.0(18.96) |
| BMI | 29.1(5.12) | 30.4(5.09) |
| | N(%) | N(%) |
| White | 105 (81.4%) | 118 (84.9%) |
| College Education or greater | 65 (50.4%) | 70 (50.4%) |
| High Blood Pressure | 61 (47.3%) | 67 (48.2%) |
| Received Chemotherapy | 67 (51.9%) | 73 (52.5%) |
| Received Radiation Therapy | 95 (73.6%) | 100 (71.9%) |
| Smoking Status | | |
| Never | 66 (51.2%) | 82 (59.0%) |
| Former | 62 (48.1%) | 55 (39.6%) |
| Current | 1 (0.7%) | 2 (1.4%) |
| Breast Cancer Stage | | |
| I | 68 (52.7%) | 64 (46.0%) |
| II | 41 (31.8%) | 49 (35.3%) |
| III | 20 (15.5%) | 26 (18.7%) |

**Table 7.** Demographics characteristics by groups of imbalanced data.

| | Intervention (N=166) mean (SD) | Control (N=167) mean (SD) |
|---|---|---|
| Age at Diagnosis | 54.3(3.18) | 65.4(4.94) |
| General Health Score | 87.3(8.48) | 57.3(13.49) |
| BMI | 25.8(1.70) | 33.5(4.49) |
| | N(%) | N(%) |
| High Cholesterol | 6 (4.7%) | 131 (94.2%) |